\documentclass{aa} 

\usepackage[colorlinks]{hyperref}
\hypersetup{ colorlinks, linkcolor=blue, citecolor=blue }

\hypersetup{
        colorlinks=true,
        allcolors=blue,
        }
\usepackage[varg]{txfonts}
\usepackage{graphicx} 
\begin{document}

\title{A robust model for the origin of optical quasi-periodic variability in supersoft X-ray sources}

\titlerunning{A robust model for the supersoft X-ray sources}
\authorrunning{W. Zhao et al.,}

\author{Weitao Zhao \inst{1,2}
\and Xiangcun Meng\inst{1,3}
\and Yingzheng Cui\inst{1,2}
\and Zheng-Wei Liu\inst{1,3}}
\institute{Yunnan Observatories, Chinese Academy of Sciences, 650216, China
\and University of Chinese Academy of Sciences, Beijing 100049, China\\e-mail: zhaoweitao@ynao.ac.cn
\and Key Laboratory for the Structure and Evolution of Celestial Objects, Chinese Academy of Sciences, Kunming 650216, China \\e-mail: xiangcunmeng@ynao.ac.cn
}


\abstract
 {Supersoft X-ray sources (SSSs) are known as possible progenitors of Type Ia supernovae. The quasi-periodic variability has been detected in the optical light curves of SSSs. However, the exact origin of such quasi-periodic observable features remains a mystery.}
 {In this paper, we aim to reproduce the observed optical quasi-periodic variability of SSSs by proposing a white dwarf (WD) accretion model with a periodic mass transfer caused by the irradiation of supersoft X-ray onto the companion star.} 
 {Assuming that a periodic mass transfer from the companion star to the WD can be caused while the supersoft X-ray irradiates the companion star, we used \textsc{MESA} to simulate the WD accretion process and the subsequent WD evolution by adopting a periodic jagged accretion rate.}
 {Comparing our results to the optical light curves of a well-observed SSS RX~J0513.9-6951, we find that our models can reproduce the quasi-periodic transition between the optical high and low states of RX~J0513.9-6951 because the periodic accretion rate can lead to the WD photosphere expands and contracts periodically in our models. In addition, we find that the transitional periods of the SSSs in our models strongly depend on the mass of the accreting WDs. The more massive the WD mass is, the shorter the transitional period.}
 {Based on our results, we suggest that the periodic mass transfer caused by the irradiation of supersoft X-ray onto the companion star may be the origin of the observed optical quasi-periodic variability in SSSs. In addition, our results indicate that the observed optical transition period of a SSS may be useful for the rough estimate of the mass of an accreting WD.}
  

\maketitle

\section{Introduction}

  Type Ia supernovae (SNe Ia) have been successfully used as the cosmological distance indicators to determine cosmological parameters \citep{Perlmutter1999,Riess1998} and thus test the evolution of the equation of the state of dark energy with time \citep{Howell2011}. However, the nature of SN Ia progenitors and their explosion mechanism remain unclear \citep{Branch1995,Hillebrandt2000,Maoz2014}. The single-degenerate (SD) model is thought to be one possible progenitor model of SNe Ia, in which a WD accretes hydrogen-rich or helium-rich materials from a nondegenerate companion star to increase its mass to the Chandrasekhar mass to trigger a SN Ia explosion \citep[e.g.,][]{Whelan1973,Nomoto1984,Meng2009a,Meng2010,Liu2018}. Supersoft X-ray sources (SSSs) have been suggested as the direct progenitor systems of SNe Ia within the SD model \citep{Nomoto1984,vandenHeuvel1992}. SSSs have extremely soft spectra, peaking at energies in the range of $15-80\,\ \rm eV$, and the total X-ray luminosity is always close to the Eddington limit ($L_{x} = 10^{36}-10^{38}\, {\rm erg \,s^{-1}}$) \citep{Greiner1991,vanTeeseling1996,Kahabka1997}. The generally accepted model for SSSs is that the accreting WDs in close binaries stably burn material from their companion star on their envelope, and this increases the mass of the WD \citep{vandenHeuvel1992}.

  RX~J0513.9-6951 is a SSS in the Large Magellanic Cloud (LMC) which was discovered in the $ROSAT$ All-Sky Survey in the 1990s \citep{Schaeidt1993}. The main observational features of RX~J0513.9-6951 are summarized as follows: (1) Its optical light curve presents a quasi-periodical transition between the optical high ($V \sim 16.6\,\mathrm{mag}$) and low states ($V \sim 17.4\,\mathrm{mag}$; \citealt{Alcock1996}). (2) The supersoft X-rays ($ \sim 30-40 \,\ \rm eV$) are only detected in the optical low state \citep{Southwell1996,Schaeidt1993,Reinsch1996}. (3) The duration of the optical high, X-ray-off state is $\sim 60 - 150$ days, and the duration of the optical low, X-ray-on state is $\sim 40$ days \citep{Cowley2002}. (4) The transition between the optical high and low state is very rapid, which is about a few days \citep{Reinsch2000}. It is believed that the transition between the optical high and low states is caused by the periodic accretion onto the WD \citep{Hachisu1996,Reinsch2000}. However, the exact origin of the periodic accretion onto the WD is still unclear.
  
 \begin{figure}
        \centering
        \includegraphics[width=0.85\columnwidth]{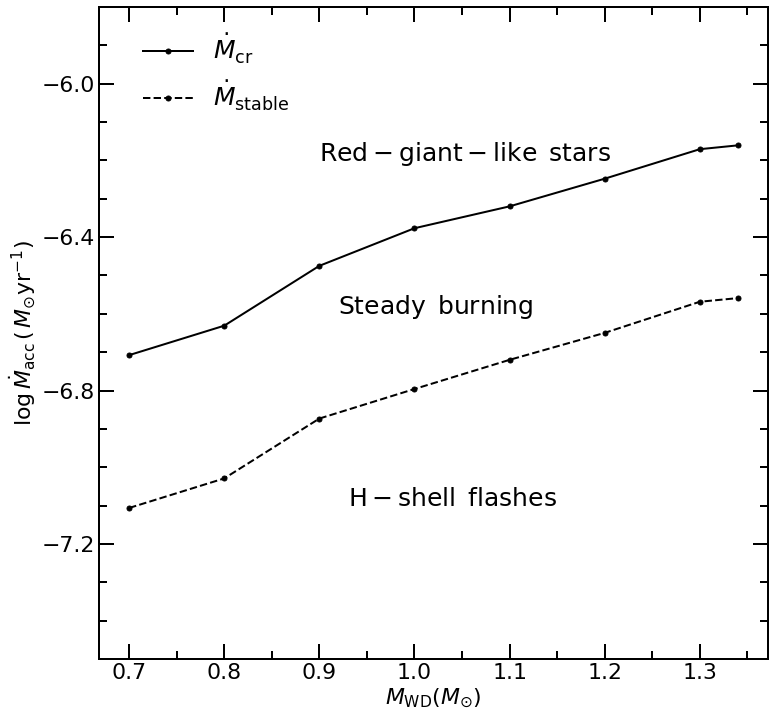}
        \caption{Possible properties of hydrogen-burning shells onto the accreting WDs in the plane of the WD mass, $M_\mathrm { WD}$, and the accretion rate, $\dot{M}_\mathrm { acc}$. Here, the initial hydrogen mass fraction was set to be $X=0.748$, and the metallicity is $Z=0.004$. The dashed and solid lines represent $\dot{M}_\mathrm { stable}$ and $\dot{M}_\mathrm { cr}$, respectively.}
        \label{fig:accretion}
 \end{figure}

 \begin{figure}
        \centering
        \includegraphics[width=0.85\columnwidth]{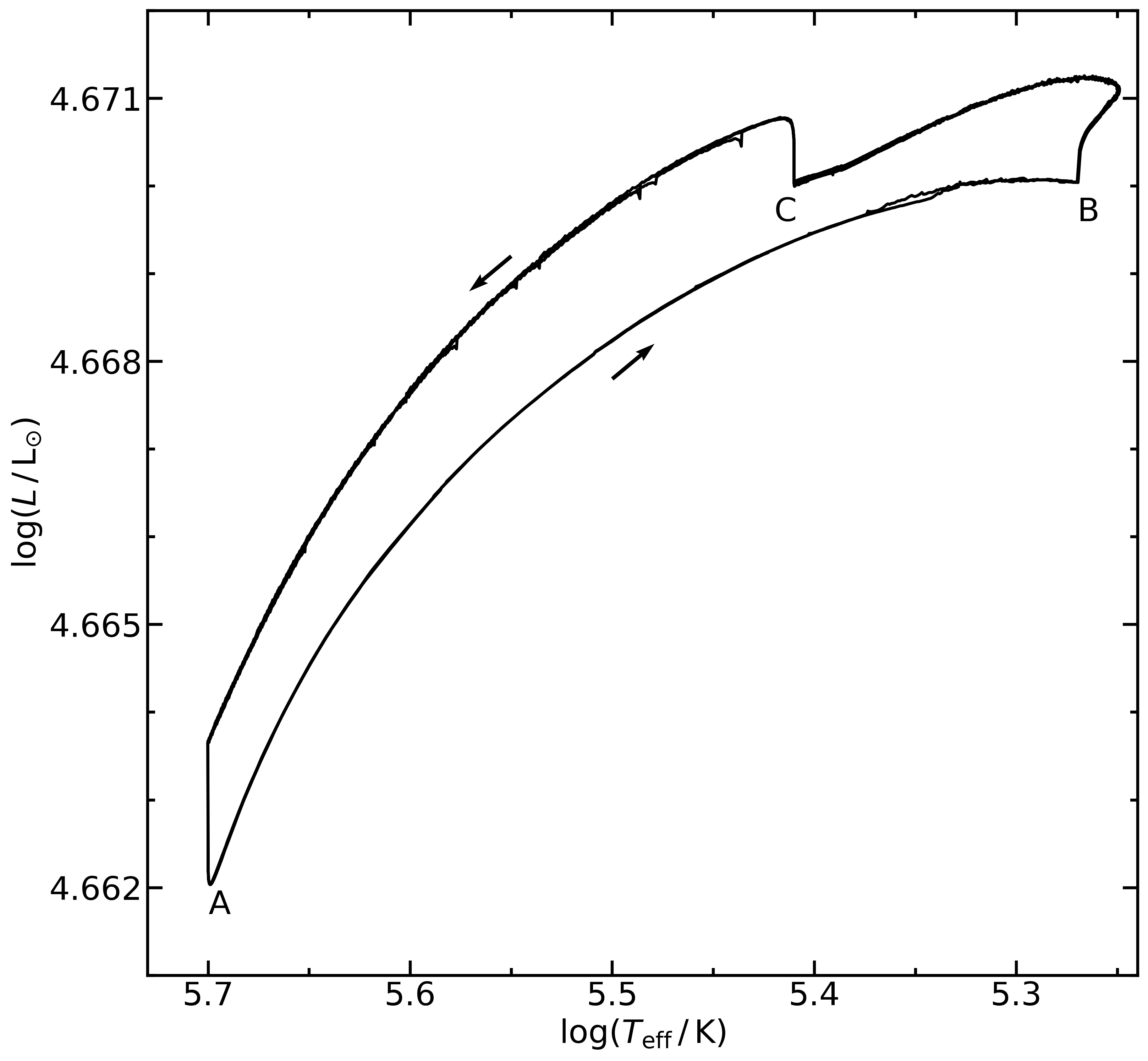}
        \caption{Evolutionary track of an accreting WD of 1.3 $M_{\odot}$ in the HR diagram. The starting point is given by "A."}
        \label{fig:HR}
\end{figure}   

  \citet{Hachisu2003c,Hachisu2003a} have suggested that the optically thick wind (hereafter, OTW) model could provide an explanation for the observed transition between the optical high and low states in SSSs. In the OTW model, the unprocessed materials would be blown away by the OTW if the accretion rate to the WD exceeds a critical accretion rate. The wind would then collide with the secondary star and strip off some masses from its surface, which would reduce the mass transfer rate and terminate the OTW. Subsequently, the mass transfer rate would increase again due to the expansion of the secondary, driving the OTW again. This periodic presence of the OTW modulates the mass transfer rate periodically, leading to a transition between the optical high and low states in the light curve of the accreting WD  \citep{Hachisu2003c,Hachisu2003a}. However, whether or not the OTW could be driven strongly depends on the abundance of Fe \citep{Iglesias1996}. The opacity is too low to drive the OTW when the metallicity $Z\le 0.002$, which means that the OTW model cannot produce SNe Ia with a redshift higher than 1.4 \citep{Kobayashi1998}. In observations, however, \citet{Rodney2015} have discovered a SN Ia with a high redshift of $z=2.26$. In addition, a high wind velocity ($\sim1000\,\rm{km \,s^{-1}}$) in the OTW model is expected to create a low-density bubble (or wind-blown cavity) around a SN Ia, which would be reflected in SN remnants (SNRs). However, the observations of seven Type Ia SNRs find no evidence for the low-density bubble or wind-blown cavity predicted by the OTW model \citep{Badenes2007}. Furthermore, the wind velocity ($\sim100 \,\rm km\,s^{-1}$) inferred from the observation of circumstellar material in SNe Ia is lower than the wind velocity of the OTW model by one order of magnitude \citep{Patat2007,Sternberg2011,Dilday2012}.
  
  Interestingly, \citet{Reinsch2000} suggested that the periodic change of the accretion disk viscosity due to the irradiation of the hot central star could lead to the change of accretion rate onto the WD. In their models, the mass transfer rates in binary systems remain unchanged. However, they did not address whether their models could reproduce the quasi-periodic transition between the optical high and low states in SSSs. 
  
  Both of the abovementioned models ignored the influence of X-ray irradiation on the companion star during the WD accretion phase. The X-ray irradiation is expected to heat the companion star and change its effective surface boundary condition, which then attenuates the mass transfer rate of the binary system \citep{PH.1991,Ritter2000}. In the present work, we assume that the supersoft X-ray irradiation onto the companion star leads to a periodical mass transfer rate. We then simulate the WD accretion process and explore whether or not such a periodical mass transfer rate could reproduce the periodic transition features observed in optical light curves of SSSs. We describe our method in Section~\ref{sec:methods}. The results of our models are presented in Section~\ref{sec:results}. We discuss our results in Section~\ref{sec:discussions}. Finally, we summarize the main conclusions in Section~\ref{sec:summary}.

\begin{figure*}
        \centering
        \includegraphics[width=1.0\textwidth]{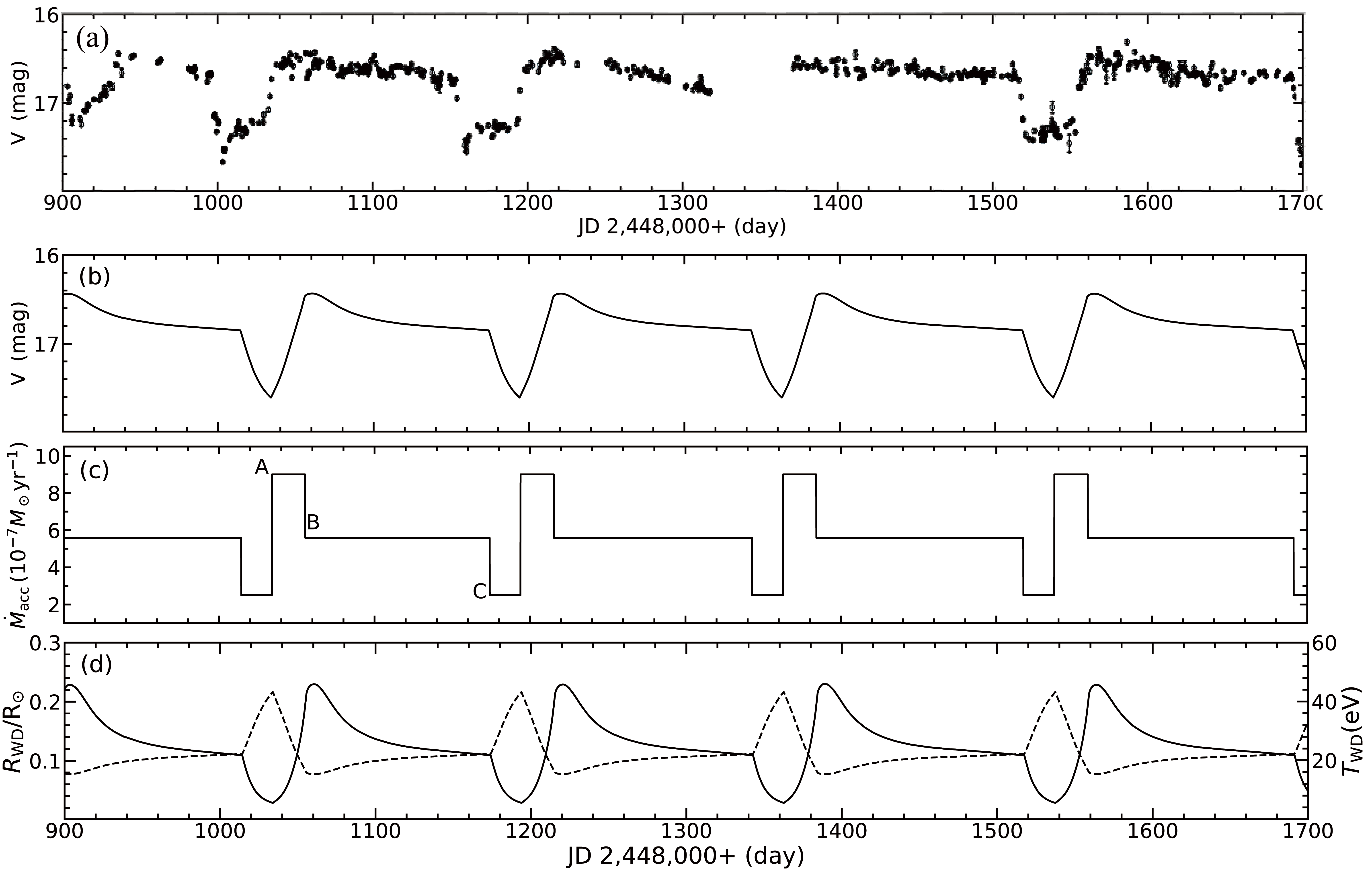}
        \caption{ Comparison between observed V-band light curve of RX~J0513.9-6951 (panel a; \citealt{Alcock1996}) and that predicted from our WD accretion model (Panel b) with an initial WD mass of $M_\mathrm { WD}$=1.3 $M_{\odot}$. Panel (c): Initial mass accretion rate ($\dot{M}_\mathrm { acc}$) adopted in our model. The positions marked by "A," "B," and "C," respectively, correspond to the accretion rates of  $\dot{M}_\mathrm { acc}^\mathrm{A}=9.0\times10^{-7} \,M_{\odot} \,{\rm yr}^{-1}$, $\dot{M}_\mathrm { acc}^\mathrm {B}=5.59\times10^{-7}\, M_{\odot}\,{\rm yr}^{-1}$, and $\dot{M}_\mathrm {acc}^\mathrm {C}=2.5\times10^{-7} \,M_{\odot} \,{\rm yr}^{-1}$.   Panel (d): Time evolution of the WD photospheric radius in units of ${R}_{\odot}$ (solid line) and the surface effective temperature of the WD in units of $\mathrm{eV}$ (dotted line) in our model. Here, we used the apparent distance modulus of $(m-M)_{V}$ = 18.7 \citep{Hachisu2003a} to calculate V-band magnitudes.}
        \label{fig:1.3main}
 \end{figure*}

\section{Methods}
\label{sec:methods}
  
  For this study, we used the Modules for Experiments in Stellar Astrophysics (MESA) code \citep[version 10398;][]{Paxton2011,Paxton2013,Paxton2015,Paxton2018,Paxton2019} to carry out 1D simulations. First, we used the suite case $make\_{co}\_{wd}$ to create the WD models with different masses, and then the $wd2$ to create the WD models with stable hydrogen burning. When the accretion rate to the WDs is within a narrow regime, that is to say when $\dot{M}_\mathrm { cr} > \dot{M}_\mathrm { acc} > \dot{M}_\mathrm { stable}$, the accreted hydrogen-rich materials are stable burning on the WDs \citep{Iben1984,Nomoto2007,Ma2013}. We set the metallicity $\mathrm {Z}$ to be 0.004, following \citet{Hachisu2003a}, and then we calculated the $\dot{M}_\mathrm { cr}$ and $\dot{M}_\mathrm {stable}$ for different WD mass following the method of \citet{Wolf2013}.
  Fig.~\ref{fig:accretion} shows two boundary curves marked by their corresponding mass-accretion rates. When the accretion rate exceeds the critical rate, $\dot{M}_\mathrm { acc} > \dot{M}_\mathrm { cr}$, we assume that there is no OTW, and the accreted materials may be piled up to form a red giant-sized envelope if the WD can continually accrete the hydrogen-rich materials \citep{Nomoto1979,Nomoto2007}; when $\dot{M}_\mathrm {acc}$ is lower than $\dot{M}_\mathrm { stable}$, the WD experiences H-shell flashes similar to nova outbursts due to unstable nuclear burning. We created different initial WD models with A mass from 1.0 to 1.3 $M_{\odot}$, and these WDs are in a stable hydrogen-burning state. The value of the hydrogen mass fraction is given by the formula $X = 0.76 - 3\,Z$ \citep{Pols1998}, that is to say the hydrogen mass fraction is 0.748 for $Z\,=\,0.004$. The ratio of mixing length to local pressure scale height, $\alpha = l/H_\mathrm{p}$, was set to be 0.4 \citep{Bergeron1995,Tremblay2015}\footnote{The inlists for our models are available at \href{https://doi.org/10.5281/zenodo.7016329}{https://doi.org/10.5281/zenodo.7016329}.}. 
  
  In this paper, we assume that the quasi-periodic X-ray irradiation from the WD onto its companion can modulate the mass transfer rate in a SSS binary system. In our future paper, we will show how X-ray irradiation leads to periodic mass transfer in detail. We periodically changed the accretion rate to the WD in order to check whether the change in the mass transfer rate can reproduce the light curve of the SSS RX J0513.9-6951. 
  
  We use Fig.~\ref{fig:HR} to describe our method in detail, where the WD mass is 1.3 $M_{\odot}$. In observations, the energy of the X-ray in the SSS RX J0513.9-6951 is between 30 $\mathrm{eV}$ and 40 $\mathrm{eV}$. Due to X-ray irradiation to the companion (point A), the mass transfer rate from the companion increases rapidly to exceed the critical accretion value of the WD. The photosphere of the WD expands rapidly, and then the effective temperature of the WD photosphere decreases gradually, which leads to the continuous attenuation of X-ray flux, and thus the mass transfer rate continues to decrease. At present, the detailed mass transfer rate with time is still unclear. To simplify the model, we used the three-stage stepped accretion rates (e.g., the accretion rate at points A, B, and C) to simulate the continuous  decreasing mass transfer rate. At point A, we assume that the accretion rate to the WD is higher than the critical accretion rate, which leads to a rapid expansion of the WD photosphere. When the effective temperature of the WD decreases below 16 $\rm eV$, we assume that the supersoft X-rays continuously attenuate, and then the mass transfer rate decreases gradually. For the accretion rate at point B, we set the mass transfer rate to be a value between $\dot{M}_\mathrm { cr}$ and $\dot{M}_\mathrm { stable}$ (point B). After the expansion for a short while, the radius of the WD photosphere decreased slowly. At the same time, the effective temperature increased. If the effective temperature was higher than 22 $\rm eV$, we decreased the mass transfer rate further to be lower than $\dot{M}_\mathrm { stable}$ (point C). Then the radius (the effective temperature) decreased (increased) quickly until the state of the WD reverted to its initial state (point A). We periodically repeated the above process to check whether or not such a process could reproduce the observed light curve of SSSs. The exact values of the accretion rate at points A, B, and C are free parameters and we also checked their effects by choosing different values. The choice of the threshold of the temperature to change the mass transfer rate was made mainly to match with the observation of RX J0513.9-6951, and we discuss the effects of this temperature choice on our results in Section~\ref{sec:discussions}.

\section{Results}
\label{sec:results}

   \begin{table*}
        \centering
        \caption{Initial parameters for different models in this paper.}
        \label{table:1}
        \begin{tabular}{ c  c  c ccc c  c  c }
        
                \hline \\
                $M_\mathrm{WD}$&  $\dot M_\mathrm { acc}^\mathrm {A}$&       $\dot M_ \mathrm{acc}^\mathrm {B}$&         $\dot M_\mathrm { acc}^\mathrm { C}$&       High&   Period& \\
                $(M_\odot)$&   $(10^{-7}\, M_\odot \,\mathrm {yr}^{-1})$&  $(10^{-7} \,M_\odot \,\mathrm {yr}^{-1})$&   $(10^{-7}\, M_\odot \,\mathrm { yr}^{-1})$&  (day)&   (day)&     \\
                \hline \\ 
                $1.3$&         12.1&                    5.0&                 0&                  18&       39           \\
                1.3&           12.1&                    5.5&                 0&                  47&       68           \\
                1.3&           12.1&                    5.5&                 2.5&                56&       85             \\
                1.3&           10.8&                    5.5&                 2.5&                58&       87           \\
                1.3&           9.0&                     5.6&                 2.5&                140&      179            \\
                1.3&           9.0&                     5.59&                2.5&                125&      163            \\
                1.3&           9.0&                     5.55&                2.5&                61&       103            \\
                1.3&           9.0&                     5.5&                 2.5&                53&       85           \\        
                1.3&           9.0&                     5.0&                 2.5&                17&       57             \\
                1.3&           8.1&                     5.5&                 2.5&                57&       105            \\
                1.3&           7.4&                     5.5&                 4&                  58&       320            \\
                1.3&           7.4&                     5.6&                 4&                  140&      402            \\
                \hline\\
                $1.2$&         10.2&                    4.3&                 0&                  610&      1418           \\
                1.2&           10.2&                    4.5&                 0&                  840&      1648           \\
                1.2&           10.2&                    4.8&                 0&                  2010&     2836           \\
                1.2&           10.2&                    4.6&                 0&                  1300&     2135           \\
                1.2&           10.2&                    4.5&                 2&                  853&      1913           \\
                1.2&           10.2&                    4.3&                 2&                  614&      1740           \\
                1.2&           9.1&                     4.5&                 2&                  852&      2038           \\
                1.2&           8.0&                     4.5&                 2&                  865&      2246           \\
                1.2&           6.8&                     4.5&                 2&                  875&      2778           \\
                1.2&           6.2&                     4.5&                 3.4&                949&      4015           \\
                1.2&           6.2&                     4.7&                 3.4&                1299&     4366           \\             
                \hline\\ 
                $1.1$&         9.2&                     4&                   0&                  4890&     8100             \\
                1.1&           9.2&                     4.3&                 1.8&                14200&    18600            \\
                1.1&           9.2&                     4.2&                 1.8&                9870&     14312            \\
                1.1&           9.2&                     4.1&                 1.8&                6050&     10556            \\
                1.1&           9.2&                     4&                   1.8&                4700&     8910             \\
                1.1&           8.2&                     4&                   1.8&                5020&     9743             \\
                1.1&           7.1&                     4&                   1.8&                4760&     10353            \\ 
                1.1&           6.1&                     4&                   1.8&                4900&     12283            \\ 
                1.1&           5.6&                     4&                   3&                  5037&     16498            \\   
                1.1&           5.6&                     4.3&                 3&                  14100&    25660            \\ 
                \hline\\                
                $1.0$&         7.6&                     3.3&                 0&                  17990&    28793             \\
                1.0&           7.6&                     3.6&                 1.4&                59000&    69300             \\
                1.0&           7.6&                     3.5&                 1.4&                35825&    50900             \\
                1.0&           7.6&                     3.4&                 1.4&                21050&    35200             \\
                1.0&           7.6&                     3.3&                 1.4&                17993&    32193             \\
                1.0&           6.7&                     3.4&                 1.4&                21000&    37003             \\
                1.0&           5.9&                     3.4&                 1.4&                21050&    39703             \\
                1.0&           5.1&                     3.4&                 1.4&                21300&    45603             \\
                1.0&           4.6&                     3.4&                 2.5&                21600&    65900             \\
                1.0&           4.6&                     3.6&                 2.5&                59100&    103330            \\
                \hline
        \end{tabular}
        \tablefoot{$M_\mathrm{WD}$ is the WD mass; $\dot M_\mathrm { acc}^\mathrm {A}$, $\dot M_ \mathrm{acc}^\mathrm {B}$, and $\dot M_\mathrm { acc}^\mathrm { C}$ are the accretion rate at point A, B, and C, respectively (see the panel (c) of Fig.~\ref{fig:1.3main}). $\mathrm{High}$ and $\mathrm{Period}$ are the duration of the optical high states and the period of the optical light curve, respectively.}
   \end{table*}

  \begin{figure*}
        \centering
        \includegraphics[width=1.0\textwidth]{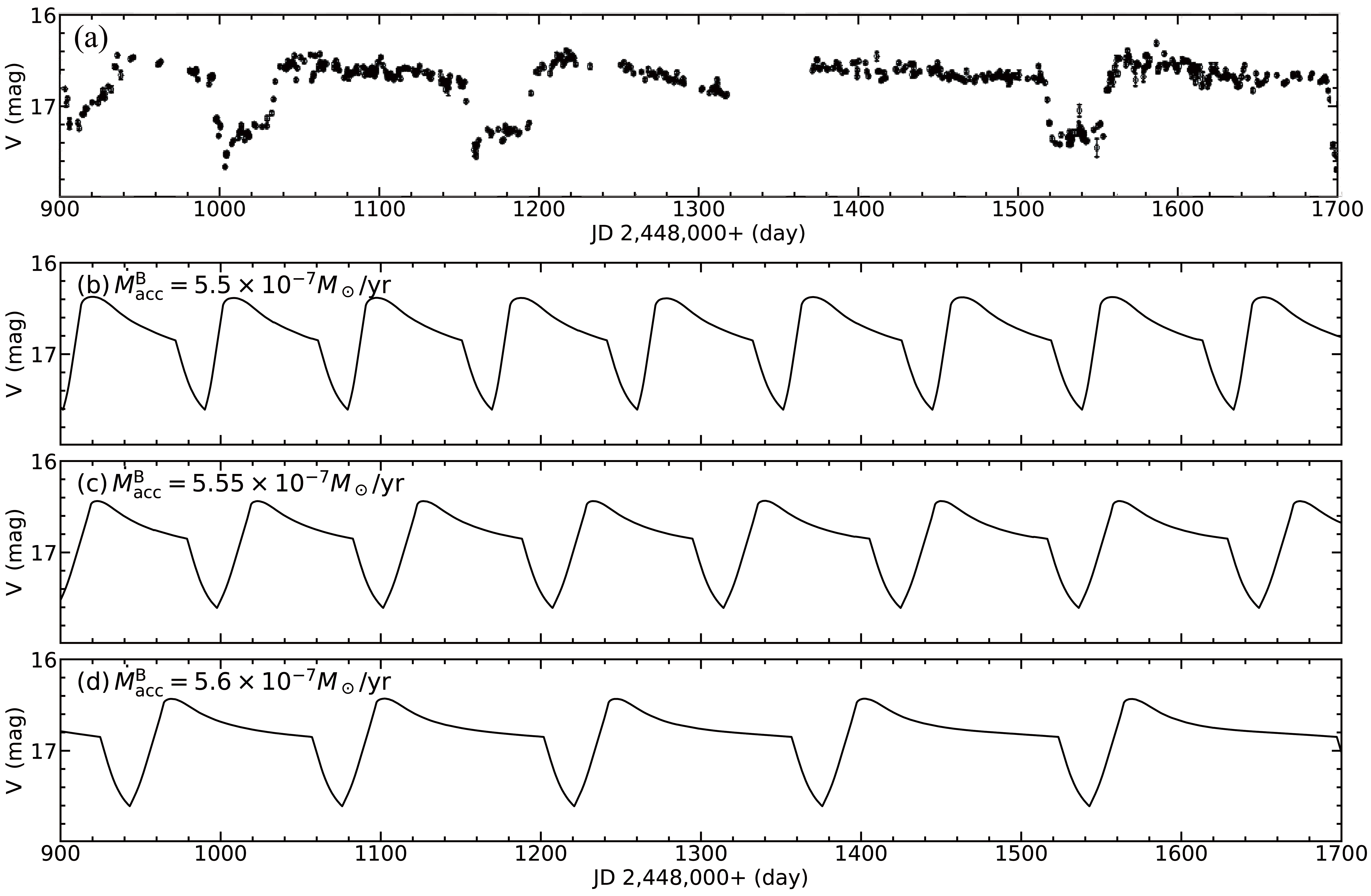}
        \caption{Evolution of V-band magnitude is similar to Fig.~\ref{fig:1.3main}, but for different accretion rates at point B (see the panel ($\mathrm{c}$) of Fig.~\ref{fig:1.3main}). Here, panels ($\mathrm{b}$), ($\mathrm{c}$), and ($\mathrm{d}$) are for the cases with $\dot{M}_\mathrm { acc}^\mathrm { B} = 5.50\times10^{-7}\,M_{\odot}\, {\rm yr}^{-1}$, $\dot{M}_\mathrm { acc}^\mathrm { B} = 5.55\times10^{-7} \, M_{\odot}\, {\rm yr}^{-1}$, and $\dot{M}_\mathrm { acc}^\mathrm {B} = 5.60\times10^{-7}\, M_{\odot}\,{\rm yr}^{-1}$, respectively.}
        \label{fig:1.3}
  \end{figure*}

  \begin{figure*}
        \centering
        \includegraphics[width=1.0\textwidth]{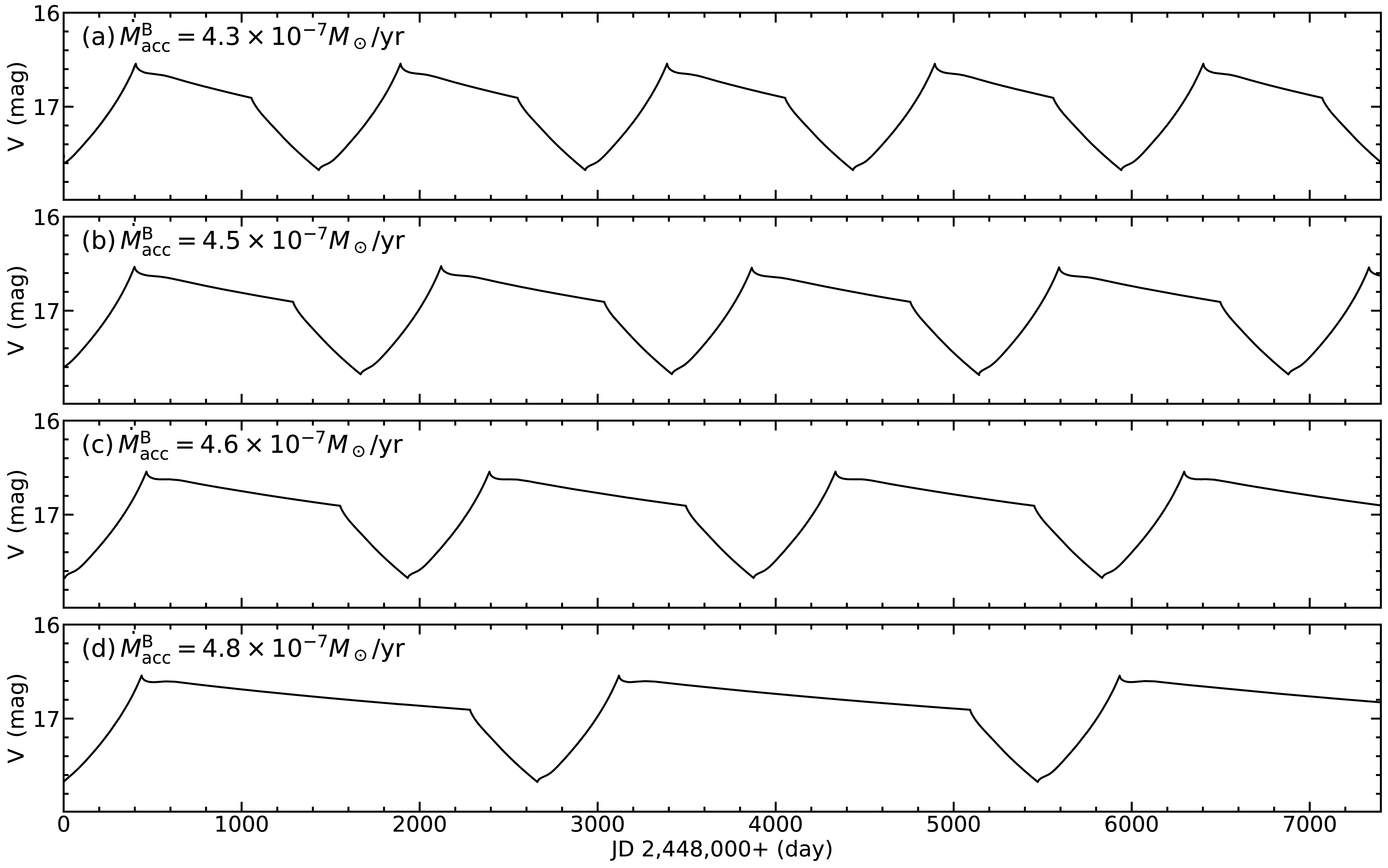}
        \caption{Similar to Fig.~\ref{fig:1.3}, but for the model with an initial WD mass of $M_\mathrm { WD}$ = $1.2 M_{\odot}$ and different accretion rates at point B (see the panel~(c) of Fig.~\ref{fig:1.3main}). Panels ($\mathrm{a}$), ($\mathrm{b}$), ($\mathrm{c}$), and ($\mathrm{d}$) show the cases with $\dot{M}_\mathrm { acc}^\mathrm { B} / \dot{M}_\mathrm { cr} = 0.75$, $\dot{M}_\mathrm { acc}^\mathrm { B} / \dot{M}_\mathrm { cr} = 0.80$, $\dot{M}_\mathrm { acc}^\mathrm { B} / \dot{M}_\mathrm { cr} = 0.82$, and $\dot{M}_\mathrm { acc}^\mathrm { B} / \dot{M}_\mathrm {cr} = 0.85$, respectively, and the exact values of $\dot{M}_\mathrm { acc}^\mathrm { B}$ are given in each panel.}
        \label{fig:1.2}
  \end{figure*}

  \begin{figure*}
        \centering
        \includegraphics[width=1.0\textwidth]{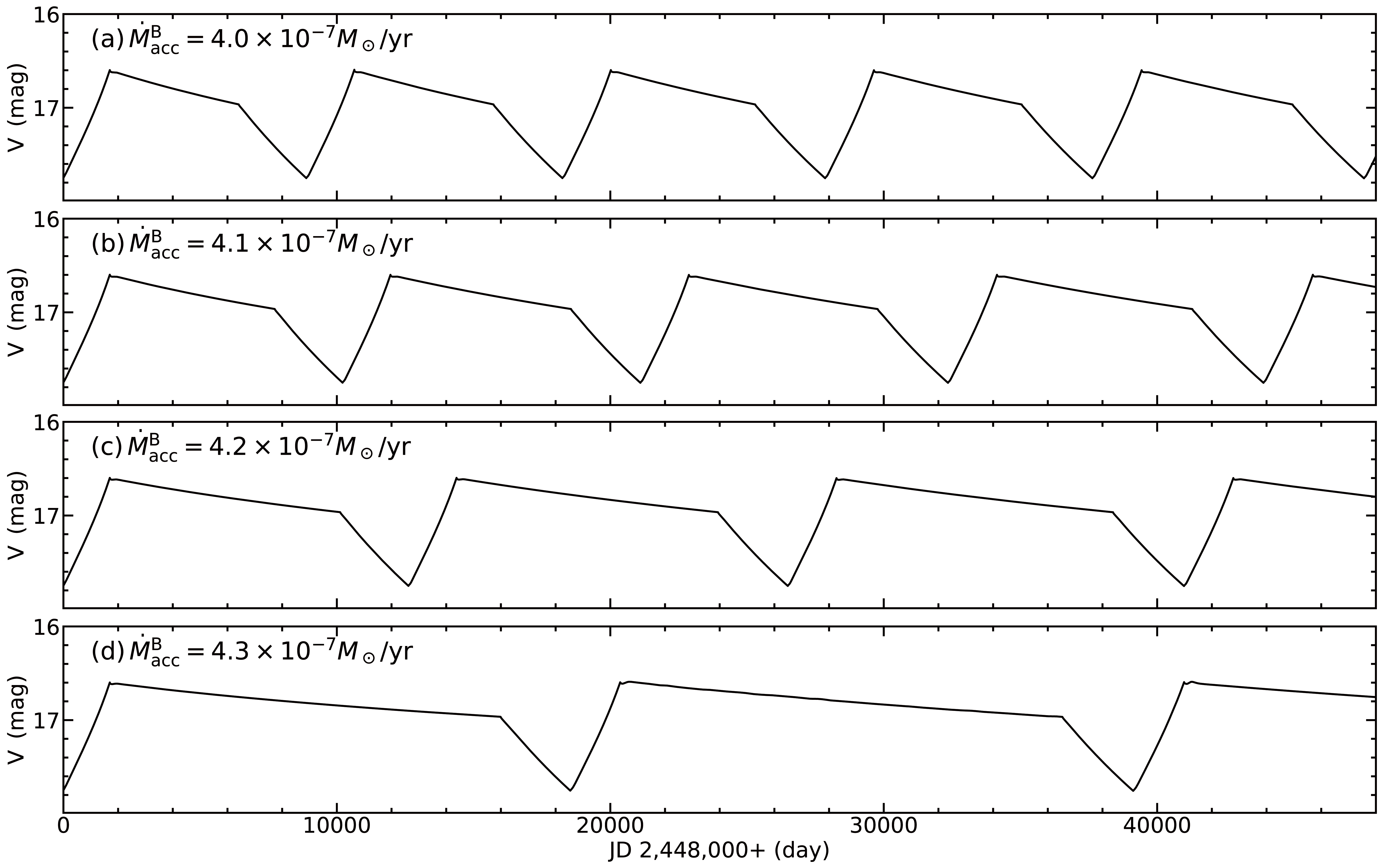}
        \caption{Similar to Fig.~\ref{fig:1.2}, but for $M_\mathrm { WD}$ = $1.1 M_{\odot}$.}
        \label{fig:1.1}
  \end{figure*}
  \begin{figure*}
        \centering
        \includegraphics[width=1.0\textwidth]{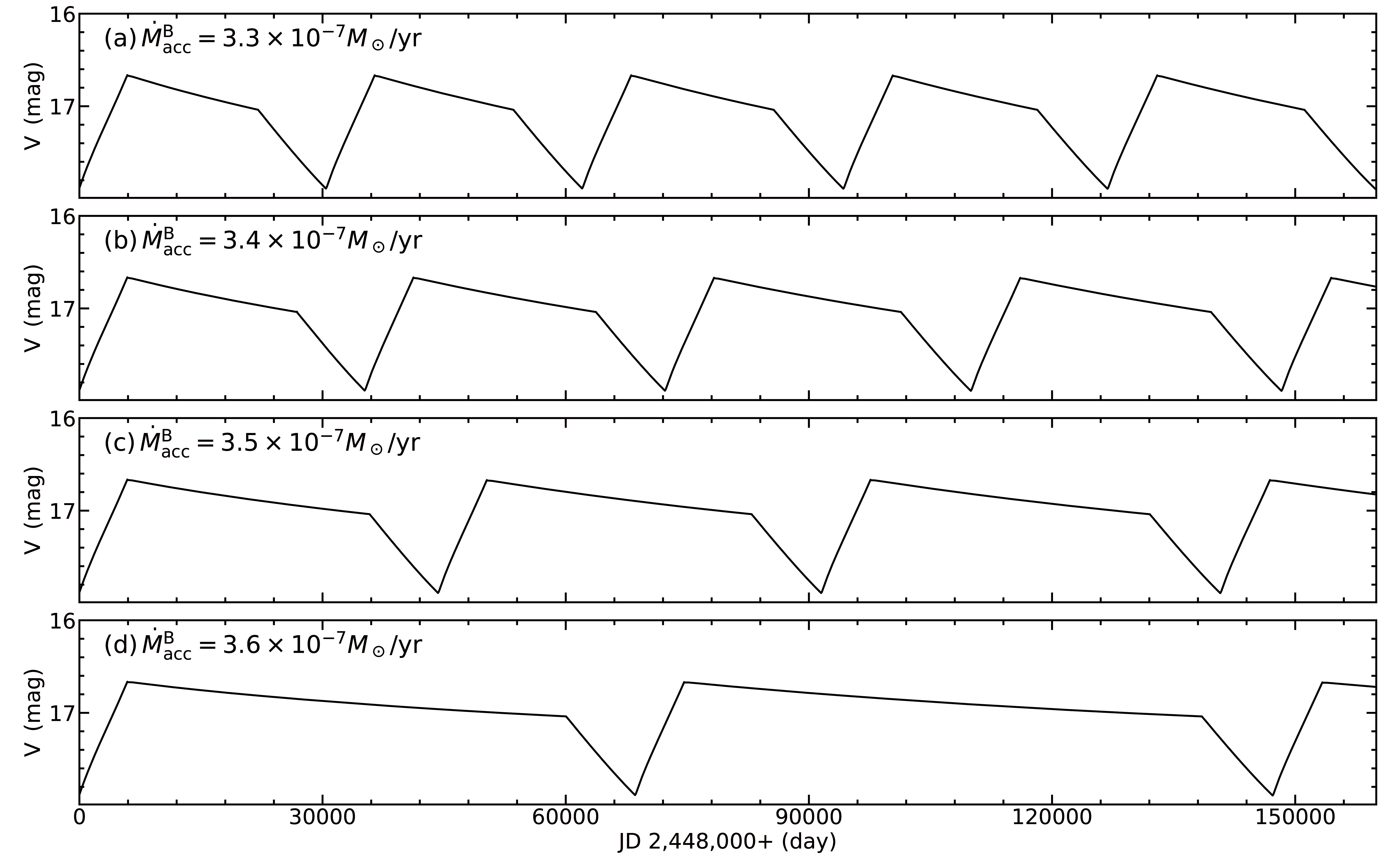}
        \caption{Similar to Fig.~\ref{fig:1.2}, but for $M_\mathrm { WD}$ = $1.0 M_{\odot}$.}
        \label{fig:1.0}
  \end{figure*}

  \subsection{A typical light curve }
  \label{3.1}
  RX J0513.9-6951 is a well-observed SSS \citep{Alcock1996,Southwell1996,Reinsch1996}. We chose a model that may match its observation as an example. The initial model is a WD of 1.3 $M_{\odot}$ with an initial envelope of $\Delta M_\mathrm { env} = 5.80 \times 10^{-7} M_{\odot}$, an initial radius of 0.028 ${R}_{\odot}$, and an initial effective temperature of 43 $\rm eV$ (point A). We assumed that the supersoft X-ray irradiates the companion star, and that the mass transfer rate increases quickly. The accretion rate at point A was set to be $\dot{M}_\mathrm { acc}^\mathrm { A} = 9.0\times10^{-7} \,{M}_{\odot} \,\mathrm { yr}^{-1}$, which is larger than $\dot{M}_\mathrm { cr}$. Then the photosphere of the WD rapidly increased to 0.214 $R_{\odot}$ within 18 days, and the effective temperature decreased to 16 $\rm eV$, where it was assumed that the WD could not radiate supersoft X-rays effectively (point B), and the mass transfer rate was expected to become low. The accretion rate at point B was set to be $\dot M_\mathrm { acc}^\mathrm { B} = 5.59\times10^{-7}\,M_{\odot}\,{\rm yr}^{-1}$, which is lower than $\dot{M}_\mathrm { cr}$, but higher than $\dot{M}_\mathrm { stable}$. The photosphere of the WD increased to 0.23 $R_{\odot}$, and then gradually decreased to 0.11 $R_{\odot}$ within 125 days. At the same time, the effective temperature increased to 22 ${\rm eV}$ (point C), where we assumed that the companion star could not recover and then reduced the mass transfer rate further to be $\dot M_\mathrm { acc}^\mathrm { C} = 2.5\times10^{-7}\, M_{\odot}\,{\rm yr}^{-1}$, which is below $\dot{M}_\mathrm { stable}$. The WD photosphere began to shrink quickly, and it took 20 days to collapse from 0.11 to 0.028 $R_{\odot}$. At the same time, the photospheric effective temperature increased quickly from 22 to 43 eV, that is the model recovered its initial condition (point A).
  
  To fit the optical light curve of RX J0513.9-6951, we adopted the apparent modulus of $ (m - M)_{V} = 18.7$ \citep{Hachisu2003a} and the color excess of $E(B - V) = 0.13$ \citep{Gansicke1998}. In Fig.~\ref{fig:1.3main}, we show the calculated V-band light curve $(\rm panel \,\ b )$, the accretion rate onto the WD $(\rm panel \,\ c )$, the radius of the WD photosphere, and the effective temperature $(\rm panel \,\ d )$, together with the MACHO observed V-band light curve of RX J0513.9-6951 $(\rm panel \,\ a )$. We simply used a blackbody spectrum to calculate the V-band light curve \citep{Liu2015}.
    
  Fig.~\ref{fig:1.3main} clearly shows that our model may reproduce most of the basic observational features of RX J0513.9-6951, that is the period and the duration of the optical high state. In our model, the light curve from point B to C represents the optical high state. However, the characteristics of optical low states are not obvious in our model. In addition, the optical transitions from high to low occur over ten days, but it is only several days in observations \citep{Reinsch2000}. These deficiencies could be mainly due to our present model being too simple. Therefore, we only focus on the duration of the optical high states and the period of the optical light curve of the SSSs in the following part of the paper. 
   
   \begin{figure}
        \centering
        \includegraphics[width=0.80\columnwidth]{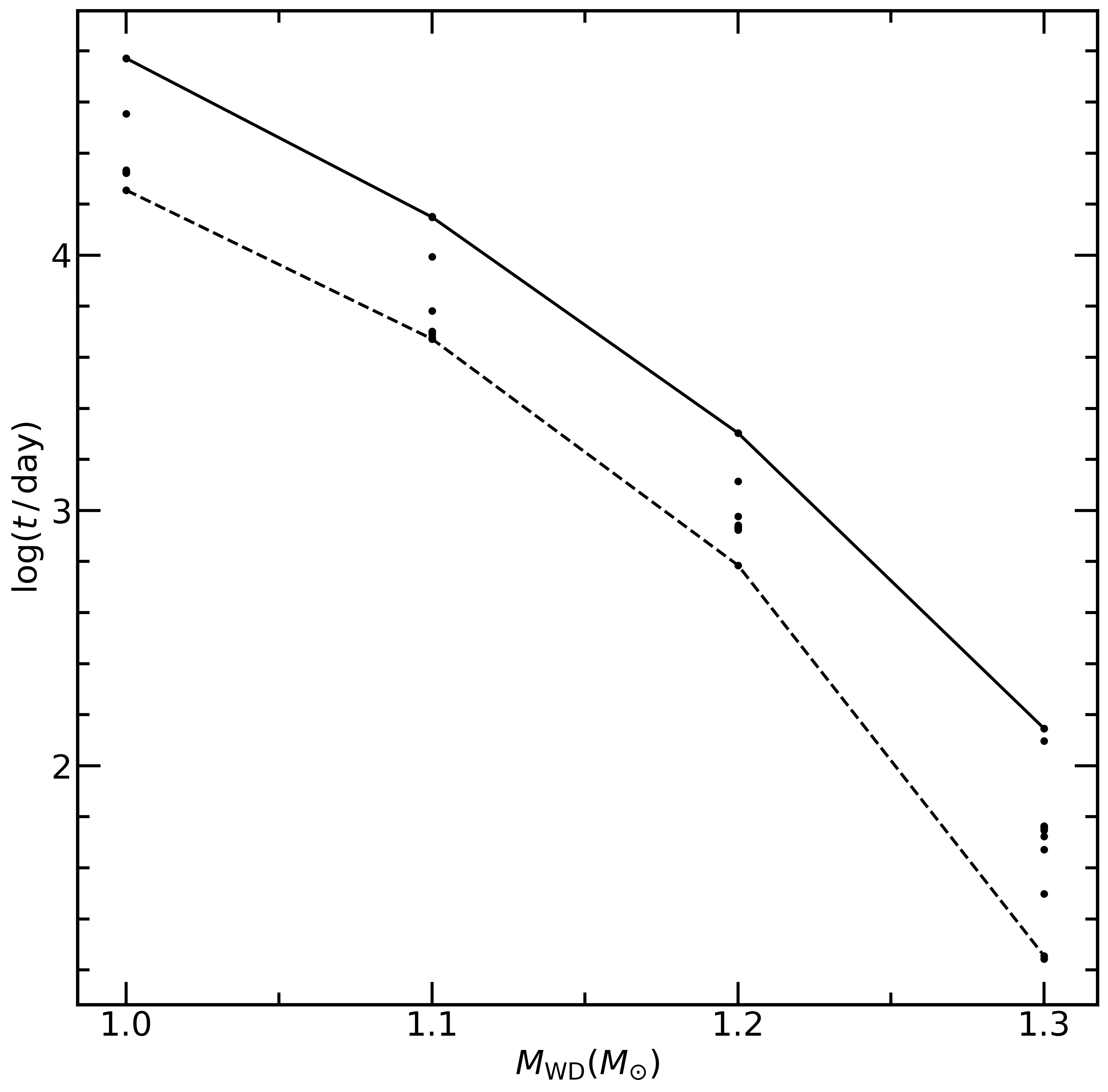}
        \caption{Duration of the optical high state ($t$) as a function of the WD mass ($M_\mathrm{WD}$) in our models. For a given WD mass, different vertical black points represent different models with various combinations of accretion rates at points A, B, and C (see Table~\ref{table:1}). The solid and dotted lines give the maximum and minimum duration boundaries for a given $M_\mathrm{WD}$, respectively.}
        \label{fig:high state and WD}
   \end{figure}

   \begin{figure}
        \centering
        \includegraphics[width=0.80\columnwidth]{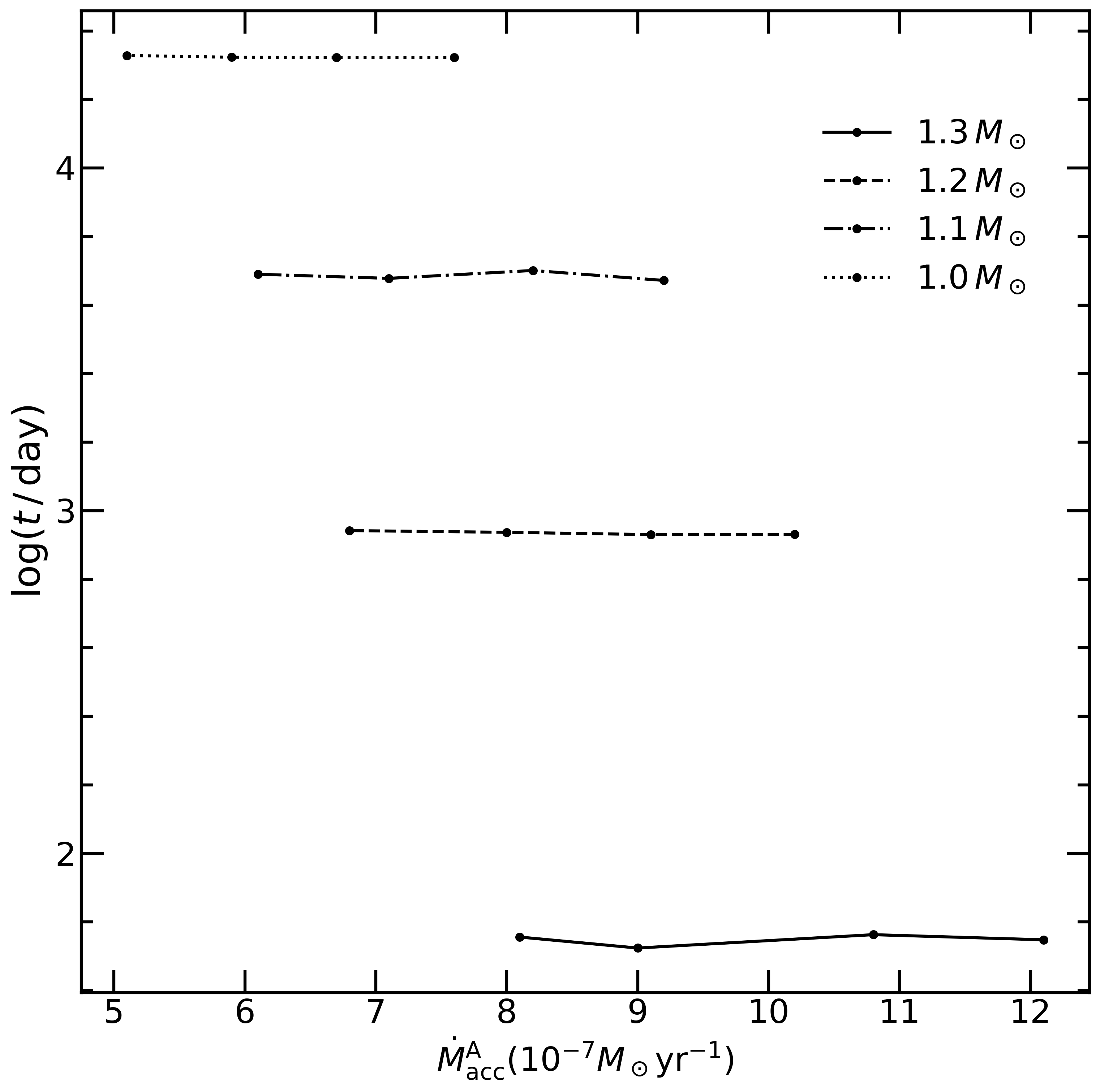}
        \caption{Duration of the optical high state as a function of the accretion rate at point A for different WD masses. We note that $\dot M_\mathrm{acc}^\mathrm{B}$ and $\dot M_\mathrm{acc}^\mathrm{C}$ were fixed for all of the models. }
        \label{fig:high state and accretion A}
   \end{figure}

   \begin{figure}
        \centering
        \includegraphics[width=0.80\columnwidth]{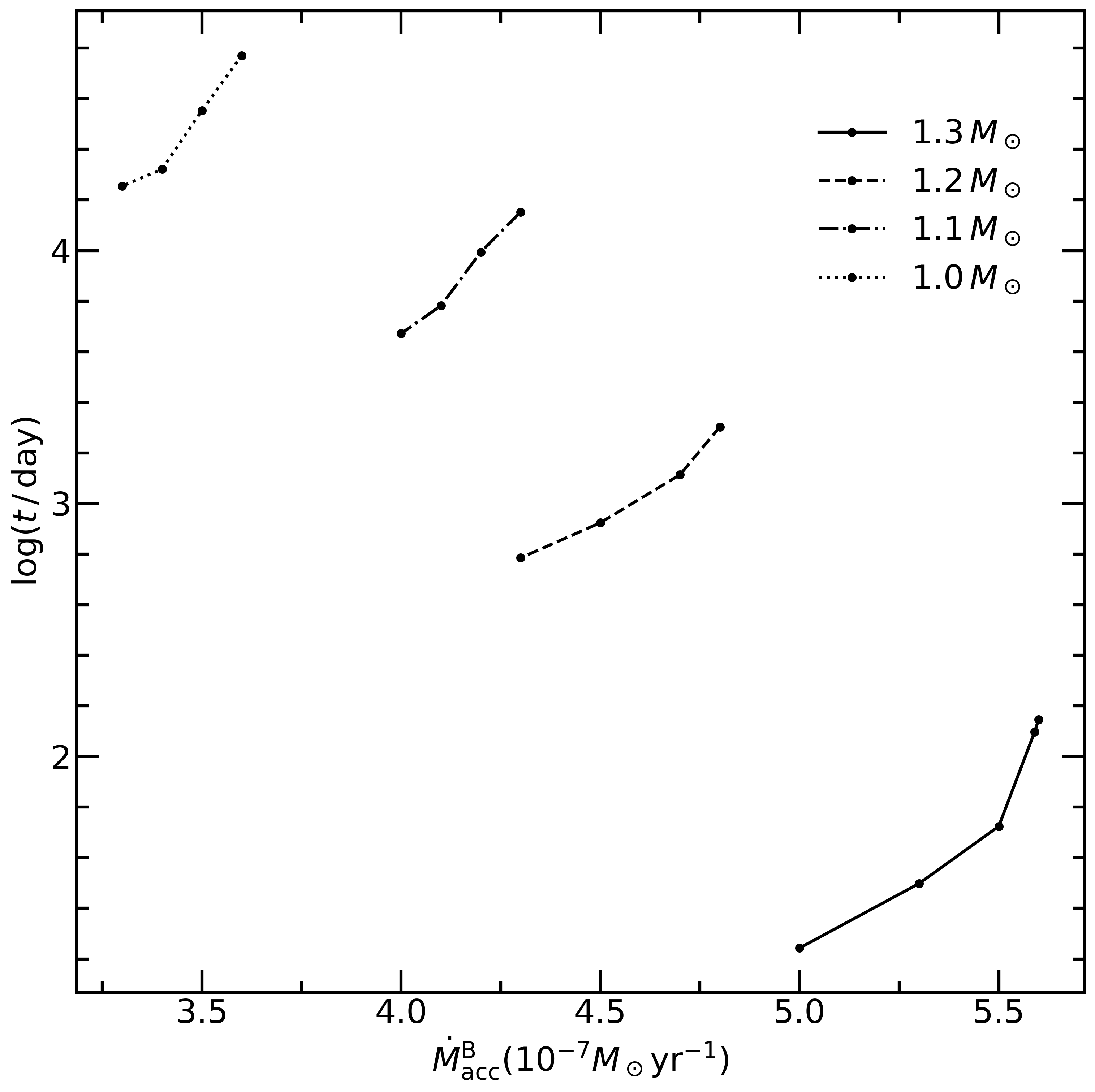}
        \caption{Similar to Fig.~\ref{fig:high state and accretion A}, but for cases with different $\dot M_\mathrm{acc}^\mathrm{B}$ for given $\dot M_\mathrm{acc}^\mathrm{A}$ and $\dot M_\mathrm{acc}^\mathrm{C} $.}
        \label{fig:high state and accretion B}
   \end{figure}

   \begin{figure}
        \centering
        \includegraphics[width=0.80\columnwidth]{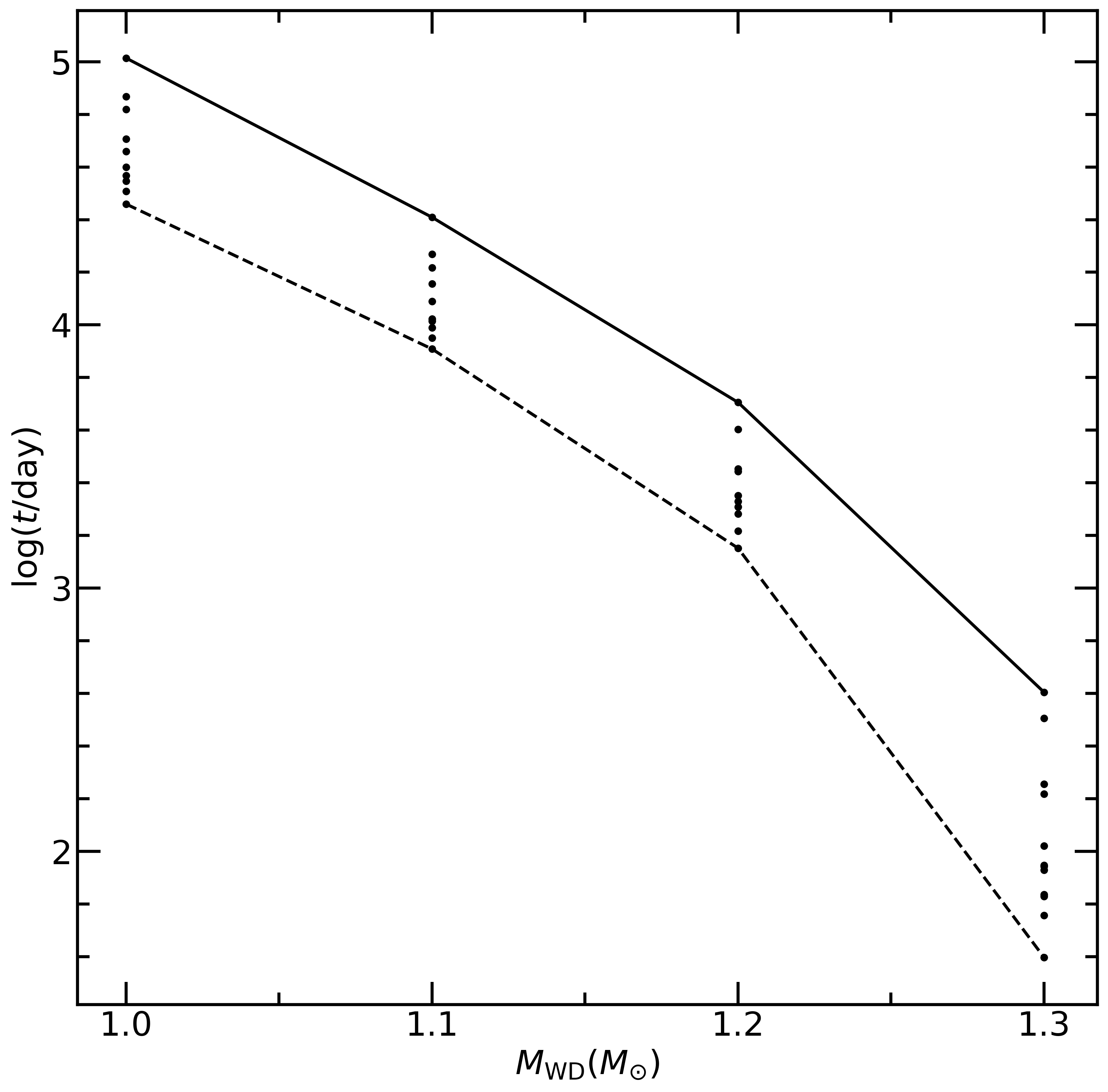}
        \caption{Similar to Fig.~\ref{fig:high state and WD}, but for the period of the optical light curve.}
        \label{fig:light curve and WD}
   \end{figure}

 \subsection{The effect of accretion rate at point B to the WD }
 \label{3.2}
   The binary parameters and the evolutionary stage of the companion determine the value of the accretion rate (mass transfer rate) \citep{Langer2000}, but we do not know the exact value of the accretion rate in RX J0513.9-6951. Here, we check the effect of different accretion rates to the WD at point B. In Fig.~\ref{fig:1.3}, we show the calculated optical light curves for different $\dot M_{\rm acc}^{\rm B}$, but given $\dot{M}_\mathrm { acc}^\mathrm { A} = 9.0\times10^{-7}\, M_{\odot}\,{\rm yr}^{-1}$ and $\dot{M}_\mathrm { acc}^\mathrm { C} = 2.5\times10^{-7}\, M_{\odot}\,{\rm yr}^{-1}$ for the model of $M_{\rm WD}$ = 1.3 $M_{\odot}$. The duration of the optical high state becomes longer for a higher $\dot M_\mathrm { acc}^\mathrm { B}$, from 53 to 140 days. In our model, the state from point B to point C corresponds to the optical high state of a SSS, in which the envelope mass of the WD, $M_\mathrm { env}$, decreases with time. The duration of the optical high state is mainly determined by the increase rate of the envelope, $\dot M_\mathrm { env}$, which can be calculated by a formula 
   \begin{equation}
    \dot{M}_\mathrm { env} = \dot{M}_\mathrm { acc} - \dot{M}_\mathrm { nuc},
  \end{equation}
   where $\dot{M}_\mathrm { nuc}$ is the hydrogen-burning rate on the WD surface. From point A to point B, the WD accumulates many materials on its surface because of $\dot{M}_\mathrm { acc}\, > \, \dot{M}_\mathrm { cr}$. When the $\dot M_\mathrm { acc}$ becomes lower than $\dot M_\mathrm { cr}$ at point B, $\dot{M}_\mathrm { nuc}$ is equal to $\dot{M}_\mathrm { cr}$ until the accumulated materials are burned out, subsequently, $M_\mathrm { env}$ gradually decreases due to the negative value of $\dot{M}_\mathrm { env}$. The higher the accretion rate of $\dot M_\mathrm { acc}^\mathrm { B}$ is, the lower the absolute value of $\dot M_\mathrm { env}$, and then the longer the duration of the optical high state. The duration of the optical high state in RX J0513.9-6951 is 60 to 150 days, which indicates that the $\dot{M}_\mathrm { acc}^\mathrm { B}$ could be (5.5 - 5.6) $\times10^{-7}\, M_{\odot}\,{\rm yr}^{-1}$.
   
  \subsection{The effect of the WD mass}
    \label{3.3}
   We also checked the effect of the WD masses, that is $M_\mathrm { WD}$\,=\, 1.2, 1.1, and 1.0 $M_{\odot}$. The results are shown in Figs.~\ref{fig:1.2}-\ref{fig:1.0}. In these WD models, we set $\dot M_\mathrm { acc}^\mathrm { A} / \dot M_\mathrm { cr} = 1.4$ and $\dot M_\mathrm { acc}^\mathrm { C} / \dot M_\mathrm { cr} = 0.3$, but different $\dot M_\mathrm { acc}^\mathrm { B} / \dot M_\mathrm { cr}$, that is $\dot M_\mathrm { acc}^\mathrm { B} / \dot M_\mathrm { cr} = $ 0.76, \,\ 0.8, \,\ 0.83, and 0.85. Regardless of the WD mass, the duration of the optical high states became longer with larger $\dot M_\mathrm { acc}^\mathrm { B}$, as did the period of the optical light curve. In addition, for given $\dot M_\mathrm { acc}^\mathrm { A} / \dot M_\mathrm { cr}$, $\dot M_\mathrm { acc}^\mathrm { B} / \dot M_\mathrm { cr}$, and $\dot M_\mathrm { acc}^\mathrm { C} / \dot M_\mathrm { cr}$, the period of the optical light curve increased with WD mass. We know that the duration of the optical high state is mainly determined by $\dot M_\mathrm {env}$. Before the hydrogen accumulated on the surface is burned out, $\dot M_\mathrm { env} = \dot{M}_\mathrm { acc}^\mathrm { B} - \dot{M}_\mathrm { cr} = k\dot{M}_\mathrm { cr} - \dot{M}_\mathrm { cr} = (k-1)\,\dot{M}_\mathrm { cr}$ can be obtained, where $k$ is a free parameter and less than 1. The lower the WD mass is, the smaller the $\dot{M}_\mathrm { cr}$, and then the smaller the absolute value of $\dot M_\mathrm { env}$, which leads to a significant increase in the duration of the optical high state.  
   
  \subsection{The optical high state}
    \label{3.4}
   We have used Figs.~\ref{fig:high state and WD}-\ref{fig:high state and accretion B} in order to clearly show the relationship between the duration of the optical high state, the WD mass, and the accretion rate. 
  
   In Fig.~\ref{fig:high state and WD}, we show the relation between the duration of the optical high state and the WD mass. This figure shows that, as indicated in Section~\ref{3.3}, the duration of the optical high state is heavily dependent on the WD mass, that is to say the more massive the WD is, the shorter the duration of the optical high state. In addition, the accretion rate to the WD also slightly affects the duration, that is for a given WD mass, the duration of the optical high state is just within a narrow range. The result is helpful to estimate the WD mass in a SSS. For example, the duration of the optical high state in RX J0513.9-6951 is from 60 to 150 days, thus the WD mass in RX J0513.9-6951 would be larger than 1.25 $M_{\odot}$, which is consistent with the previous estimation \citep{Hachisu2003a}.
   
   In Fig.~\ref{fig:high state and accretion A}, we show the relation between the duration of the optical high state and the accretion rate at point A, but with given $\dot{M}_\mathrm { acc}^\mathrm { B}$ and $\dot{M}_\mathrm { acc}^\mathrm { C}$ for a given WD mass. For a given WD mass, the duration of the optical high states hardly varies with different $\dot{M}_\mathrm { acc}^\mathrm { A}$. In fact, as discussed in Section~\ref{3.2}, for a given $\dot{M}^\mathrm { B}_\mathrm { acc}$, the duration of the optical high state is mainly determined by how many materials are accumulated on the surface of the WD from point A to B. For a given WD mass, the mass of accumulated materials from point A to B is a constant, therefore, the duration of the optical high state does not depend on the accretion rate at point A.
   Again, in Fig.~\ref{fig:high state and accretion A}, for a given $\dot{M}_\mathrm { acc}^\mathrm { A}$, the duration of the optical high state significantly decreases with the WD mass.
   
   In Fig.~\ref{fig:high state and accretion B}, we show the duration of optical high states versus $\dot{M}_\mathrm { acc}^\mathrm { B}$, but we give $\dot{M}_\mathrm { acc}^\mathrm { A}$ and $\dot{M}_\mathrm { acc}^\mathrm { C}$ for different WD masses. This result is similar to the result of Section~\ref{3.2}, that is for a given WD mass, the duration of the optical high states increases with $\dot M_\mathrm { acc}^\mathrm { B}$.
   
  \subsection{The period of the optical light curve}
    \label{3.5}
   In Fig.~\ref{fig:light curve and WD}, we present that the period of the optical light curve versus the WD mass. Similar to Fig.~\ref{fig:high state and WD}, Fig.~\ref{fig:light curve and WD} shows that the period of the optical light curve is heavily dependent on the WD mass, that is the more massive the WD is, the shorter the period of the optical light curve. In addition, the accretion rate to the WD also slightly affects the period, in other words for a given WD mass, the period of the optical light curve is within a narrow range. The results may also be helpful to estimate the WD mass in a SSS. Such as in RX J0513.9-6951, the period of the optical light curve is from 100 to 190 days, thus its WD mass may be larger than 1.25 $M_{\odot}$, which is consistent with the estimation in Section~\ref{3.4}. In addition, the result also indicates that if a WD in a SSS is less massive than 1.2\,$M_{\odot}$, it becomes difficult to observe its periodic optical light curve from the expansion and contraction of its photosphere because its period is much longer than 1,000 days.
   
\section{Discussion} 
\label{sec:discussions}

    \begin{figure}
        \centering
        \includegraphics[width=0.80\columnwidth]{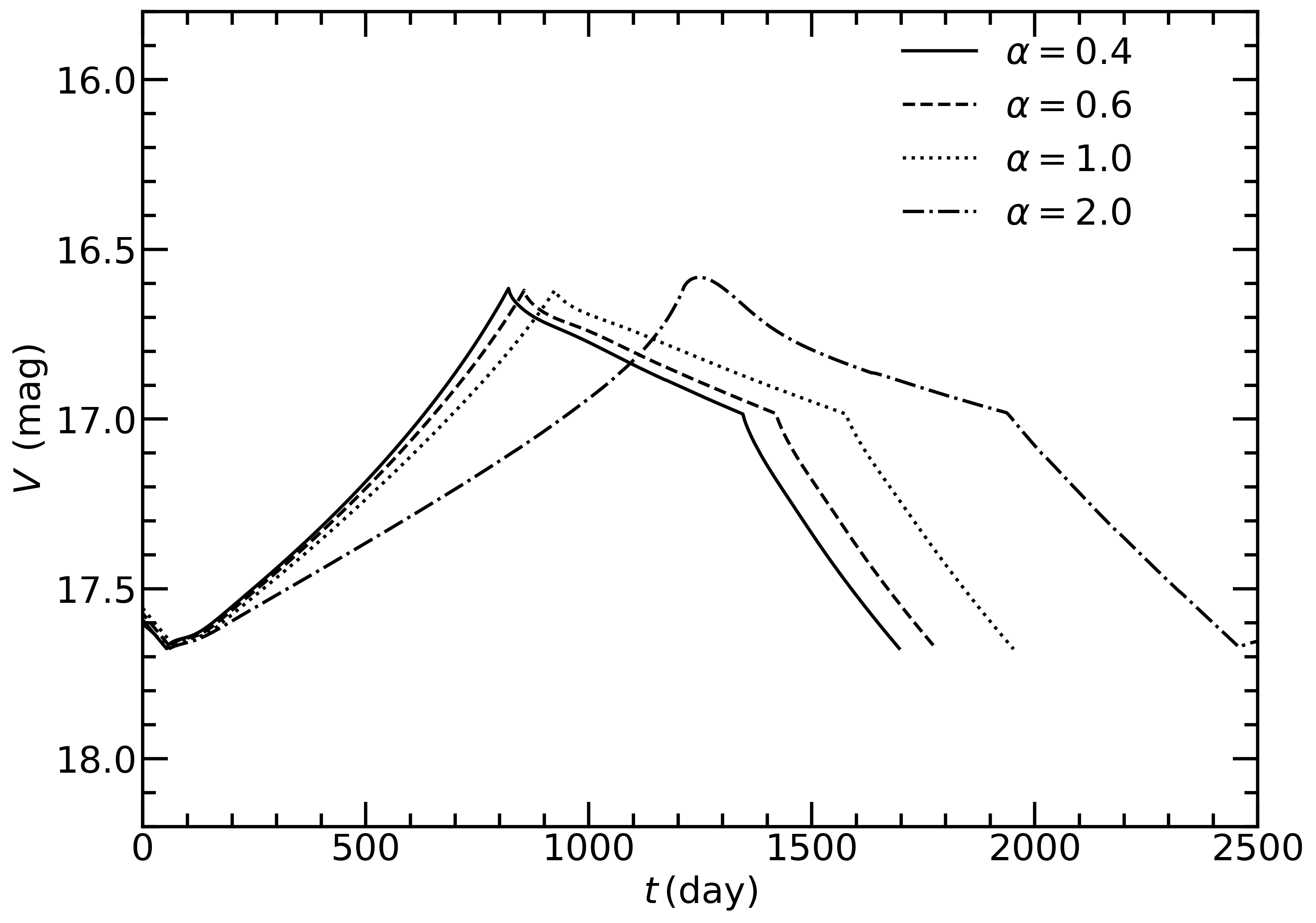}
        \caption{Predicted V-band light curves of our models for different mixing-length parameters, i.e., $\alpha=0.4$ (solid line), $0.6$ (dashed line), $1.0$ (dotted line), and $2.0$ (dot-dashed line), respectively. Here, the initial WD mass is $M_\mathrm{WD}$ = 1.2 $M_{\odot}$, and we set $\dot{M}_\mathrm { acc}^\mathrm { A} = 8.0\times10^{-7}\, M_{\odot}\,{\rm yr}^{-1}$, $\dot{M}_\mathrm { acc}^\mathrm { B} = 4.0\times10^{-7}\, M_{\odot}\,{\rm yr}^{-1}$, and $\dot{M}_\mathrm { acc}^\mathrm { C} = 0$.}
        \label{fig:alpha}
   \end{figure} 
   
   \begin{figure}
        \centering
        \includegraphics[width=0.80\columnwidth]{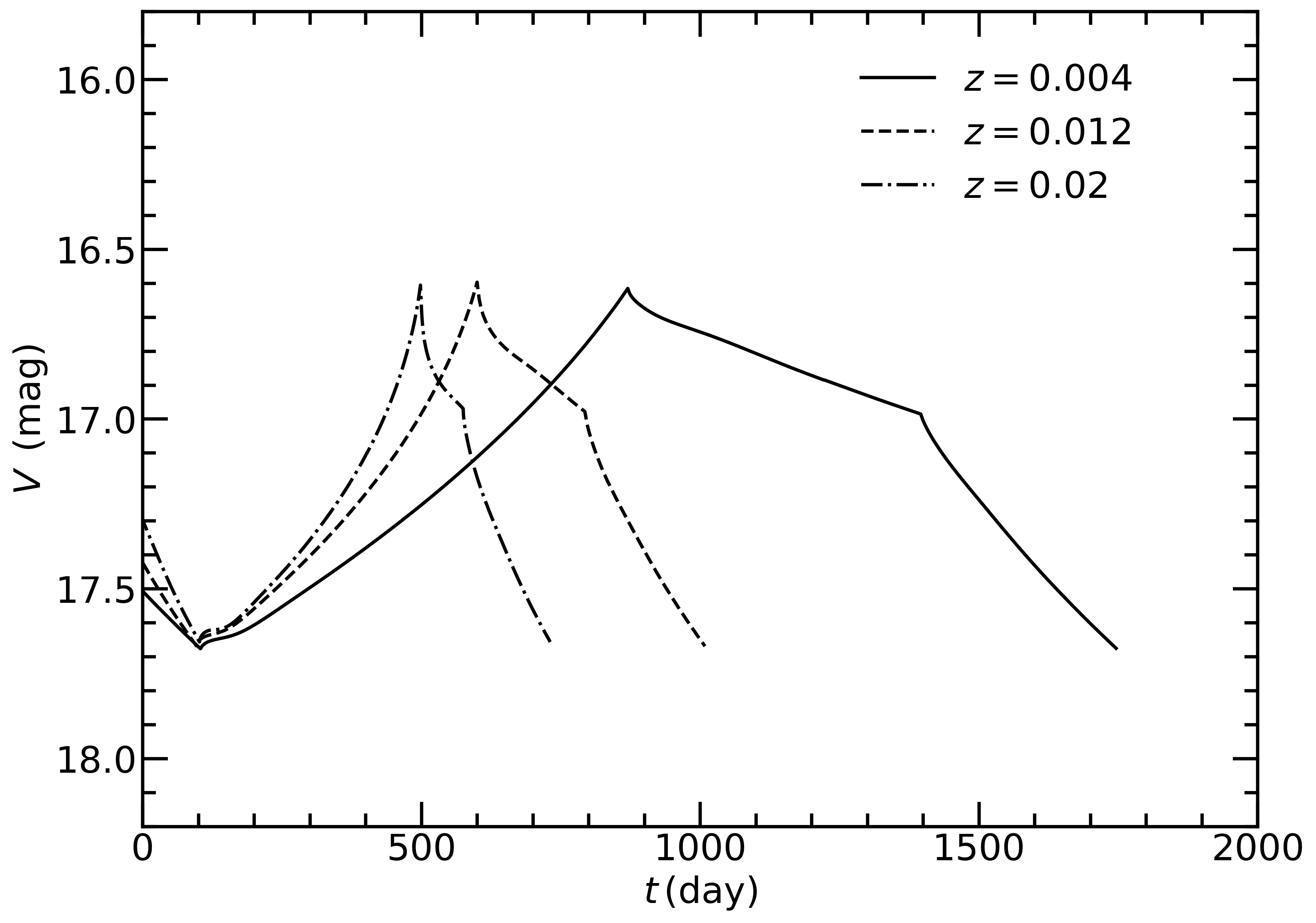}
        \caption{Similar to Fig.~\ref{fig:alpha}, but for different metallicity, i.e., $Z = 0.004$ (solid line), $0.012$ (dashed line), and $0.02$ (dot-dashed line), respectively.}
        \label{fig:z}
   \end{figure} 
   
   \begin{figure}
        \centering
        \includegraphics[width=0.80\columnwidth]{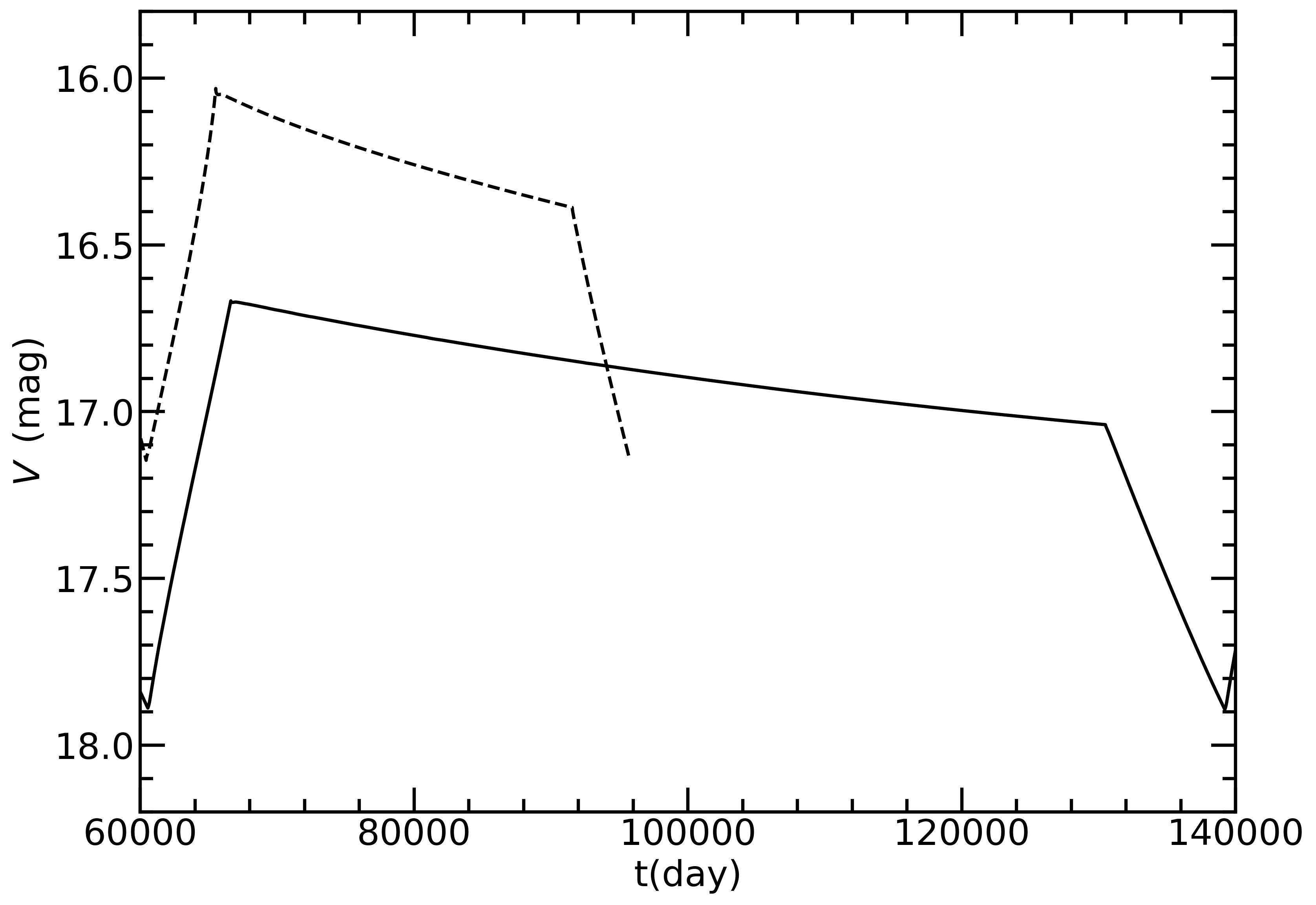}
        \caption{Predicted V-band light curves of our models for different temperature criteria, where $M_\mathrm{WD}$ = 1.0 $M_{\odot}$, $\dot{M}_\mathrm { acc}^\mathrm { A} / \dot{M}_\mathrm { cr} = 1.8$, $\dot{M}_\mathrm { acc}^\mathrm { B} / \dot{M}_\mathrm { cr} = 0.8$, and $\dot{M}_\mathrm { acc}^\mathrm { C} / \dot{M}_\mathrm { cr} = 0.3$. The solid and dashed lines represent the light curve for the temperature criterion from the massive WD and the low-mass WD, respectively.}
        \label{fig:Vmag}
   \end{figure}

   It has been suggested that the transition between the optical high (X-ray-off) and low (X-ray-on) states is derived from the contraction or expansion of the WD photosphere resulting from the variation of the accretion rates to the WD.\ However, the origin of the variation of the accretion rates is still unclear \citep{Pakull1993,Reinsch1996,Reinsch2000,Southwell1996,Meyer-Hofmeister1997}. In this paper, we assume that periodically supersoft X-rays from a WD irradiate its companion star, which leads to the expansion and contraction of the companion star, and then a periodical mass transfer rate. Our model can reproduce the main observational properties of the optical light curve of SSS RX J0513.9-6951, which indicates that the periodic mass transfer caused by the irradiation of supersoft X-rays onto the companion may be the origin of the observed optical quasi-periodic variability in SSSs.
 
   In this paper, we set $\alpha = 0.4$. For a DA WD, \citet{Bergeron1995} found that models calculated with $\alpha = 0.6$ may match the observations very well. However, the effective temperatures of the WDs are generally lower than $34,000\, \mathrm{K}$ \citep{Cukanovaite2019,Tremblay2015}. In this paper, the effective temperatures of the accreting WDs are more than 100,000 K, where the value of $\alpha$ is still unclear. Considering that the WDs in SSSs have a very high accretion rate and possible X-ray irradiation feedback, we tested the effects of different mixing length parameters. In Fig.~\ref{fig:alpha}, we show the effects of different $\alpha$ on the optical light curve for the model of $M_\mathrm{WD}$ = 1.2 $M_{\odot}$. The solid, dashed, dotted, and dot-dashed lines are the optical light curve from $\alpha = 0.4$, $0.6$, $1.0$, and $2.0$, respectively, and their periods for the optical light curve are 1650, 1680, 1980, and 2380 days, respectively.
   Hence, the effect of increasing $\alpha$ is similar to that of increasing $\dot{M}_\mathrm { acc}^\mathrm {B}$, but the effect is much less significant than that of $M_\mathrm { WD}$. As a result, the uncertainty from different $\alpha$ cannot significantly affect our conclusion. 
   
  The LMC metallicity is one-third of the solar metallicity \citep{Hachisu2003a}. Here, we assume that the metallicity is 0.004 for the SSS RX J0513.9-6951. However, other SSSs may have different metallicities from these in LMC, and thus we also tested the effects of different metallicities on the optical light curve. In Fig.~\ref{fig:z}, we show the effects of the different metallicity $Z$ on the optical light curve for the model of $M_\mathrm{WD}$ = 1.2 $M_{\odot}$. The solid, dashed, and dot-dashed lines are the optical light curves from $Z = 0.004$, 0.012, and 0.02, respectively, and the periods of the optical light curves are about 1650, 900, and 620 days, respectively. The period of the optical light curve decreases with the metallicity. However, the effect of increasing $Z$ is similar to that of decreasing $\dot{M}_\mathrm { acc}^\mathrm {B}$, and the effect of metallicity is also much less significant than that of $M_\mathrm { WD}$.
   
   The effects of irradiation on the evolution of low-mass X-ray binaries (LMXBs) have been considered in many works in which the X-rays irradiate the companion stars, which drives a higher mass transfer \citep{PH.1991,Ritter1994,Ritter2000,Harpaz1991,Harpaz1994,Hameury1993}. \citet{King1995} studied the irradiation of the secondary star that can drive mass transfer onto the primary star in cataclysmic variables (CVs). Even though the secondary star is a low-mass main-sequence companion star, it can provide a very high mass transfer rate, which much exceeds the stable hydrogen-burning rate of the WD \citep{Ritter2000,Harpaz1991,Harpaz1994}. Due to their longer orbital periods, the secondary stars in SSSs are generally more massive than those in LMXBs. Because the X-rays have a very high luminosity and burst periodically in SSSs, we assume that the mechanism of X-ray irradiation on the companion may also provide a periodic mass transfer rate. In this paper, we have adopted a periodic jagged accretion rate to simulate the accretion rate onto the WD. However, when a WD accretes materials from its companion, the accretion rate to the WD with time is continuous, which could have an impact on the duration of the optical high and low states. In future work, we will examine whether or not or how the irradiation of the supersoft X-rays can modulate the periodic mass transfer rate in detail.
   
   In Section~\ref{sec:methods}, we set that the change of the accretion rate to the WD which was determined by the surface effective temperature of the WD, and the choice of this temperature criterion was based on the blackbody radiation temperature range of the SSS given by the observation of RX J0513.9-6951. For the low-mass WDs, its surface effective temperature in a stable hydrogen-burning state is lower than that of a massive WD, that is it may need smaller temperature criteria than the low-mass WDs. However, the temperature criterion of the low-mass WDs is still unclear because of the lack of observational data of SSSs with a low-mass WD and an optical quasi-periodic variability. In order to check the influence of the different temperature criteria on our results, we simply chose the model of $M_\mathrm{WD}$ = 1.0 $M_{\odot}$ as a low-mass example and the model of $M_\mathrm{WD}$ = 1.3 $M_{\odot}$ as a high-mass example. We assumed that the value of temperature criterion at three points from the low-mass WD to be half of that from the massive WD. Fig.~\ref{fig:Vmag} shows the calculated V magnitude from the different temperature criteria. The solid and dashed lines represent the light curve for the temperature criterion from the massive WD and the low-mass WD, respectively. The period of the optical light curve from the temperature criterion of the low-mass WD is about two times shorter than that from the massive WD. A lower temperature criterion means that the situation for changing the accretion rate in our models is easier to achieve, and thus a shorter period.
   In addition, the magnitude of the temperature criterion from the low-mass WD is $0.7\,\mathrm{mag}$ higher than that from the massive WD. This is because a less massive WD has a lower effective temperature, and thus more photo energy is radiated in the optical band. Although the period of the light curve becomes longer with the temperature criterion, the effect is also significantly less than that of $M_\mathrm { WD}$. Therefore the temperature criterion cannot significantly affect our conclusion, that is the more massive the WD is, the shorter the period of the optical light curve. 
   
   In this paper, we have assumed that when the accretion rate exceeds the critical accretion rate, the accreting WD expands, instead of the occurrence of the OTW. Whether the OTW may occur or not is still an open problem \citep[see the discussion in][]{Meng2017}. For example, if the WD may continuously accrete materials from its companion, the dynamical simulation shows that the OTW may not occur \citep{Cui2022}.
   
\section{Summary} 
\label{sec:summary}

    In this paper, we have assumed that the irradiation of supersoft X-ray from the accreting WD onto its companion star could lead to a periodic mass transfer rate of the binary system. We used \textsc{MESA} \citep{Paxton2011,Paxton2013,Paxton2015,Paxton2018,Paxton2019} to model the WD accretion process and the subsequent WD evolution by adopting a periodic jagged accretion rate. We find that the periodic mass transfer results in the periodic contraction or expansion of the WD photosphere, leading the light curve of the accreting WD to present optical quasi-periodic variability. Comparing our results to the observations of a SSS RX~J0513.9-6951, our model can reproduce the quasi-periodical transition properties in the optical light curve of RX~J0513.9-6951. This suggests that the periodic mass transfer caused by the irradiation of supersoft X-ray onto the companion may be the origin of the observed optical quasi-periodic variability in SSSs. In addition, we find that the transition period of the optical light curve in our models is strongly dependent on the WD mass. The more massive the WD is, the shorter the transitional period. This provides a way to roughly estimate the WD mass in SSSs based on the observed transitional period. For example, we can estimate that the WD mass of RX J0513.9-6951 could be more massive than 1.25 $M_{\odot}$ because it has the period of the optical light curve from 100 to 190 days. Moreover, our results indicate that it would be difficult to observe the periodic optical variability in SSSs if their WD masses are less massive than 1.2\,$M_{\odot}$ because the transition periods of such SSSs are much longer than $1,000 \,\mathrm{days}$.

\begin{acknowledgements}
   This work was supported by the National Key R\&D Program of China (No. 2021YFA1600401) and the Natural Science Foundation of China (nos. 11973080 and 11733008, 11873016). We acknowledge science research grants from the China Manned Space Project, no. CMS- CSST-2021-B07. XM acknowledges the support by the Yunnan Ten Thousand Talents Plan-Young Elite Talents Project, and CAS "Light of West China" Program. ZWL was supported by the National Key R\&D Program of China (No. 2021YFA1600401), the Chinese Academy of Sciences (CAS) and the Natural Science Foundation of Yunnan Province (No.\ 202001AW070007).

\end{acknowledgements} 
	\nocite{*}
	\bibliography{WD_accretion.bbl}

\begin{thebibliography}{69}
\expandafter\ifx\csname natexlab\endcsname\relax\def\natexlab#1{#1}\fi

\bibitem[{{Alcock} {et~al.}(1996){Alcock}, {Allsman}, {Alves}, {Axelrod},
  {Bennett}, {Charles}, {Cook}, {Freeman}, {Griest}, {Guern}, {Lehner},
  {Livio}, {Marshall}, {Peterson}, {Pratt}, {Quinn}, {Rodgers}, {Southwell},
  {Stubbs}, {Sutherland}, \& {Welch}}]{Alcock1996}
{Alcock}, C., {Allsman}, R.~A., {Alves}, D., {et~al.} 1996, \mnras, 280, L49

\bibitem[{{Badenes} {et~al.}(2007){Badenes}, {Hughes}, {Bravo}, \&
  {Langer}}]{Badenes2007}
{Badenes}, C., {Hughes}, J.~P., {Bravo}, E., \& {Langer}, N. 2007, \apj, 662,
  472

\bibitem[{{Bergeron} {et~al.}(1995){Bergeron}, {Wesemael}, {Lamontagne},
  {Fontaine}, {Saffer}, \& {Allard}}]{Bergeron1995}
{Bergeron}, P., {Wesemael}, F., {Lamontagne}, R., {et~al.} 1995, \apj, 449, 258

\bibitem[{{B{\"o}hm-Vitense}(1958)}]{Bohm-Vitense1958}
{B{\"o}hm-Vitense}, E. 1958, \zap, 46, 108

\bibitem[{{Borkowski} {et~al.}(2006){Borkowski}, {Williams}, {Reynolds},
  {Blair}, {Ghavamian}, {Sankrit}, {Hendrick}, {Long}, {Raymond}, {Smith},
  {Points}, \& {Winkler}}]{Borkowski2006}
{Borkowski}, K.~J., {Williams}, B.~J., {Reynolds}, S.~P., {et~al.} 2006, \apjl,
  642, L141

\bibitem[{{Branch} {et~al.}(1995){Branch}, {Livio}, {Yungelson}, {Boffi}, \&
  {Baron}}]{Branch1995}
{Branch}, D., {Livio}, M., {Yungelson}, L.~R., {Boffi}, F.~R., \& {Baron}, E.
  1995, \pasp, 107, 1019

\bibitem[{{Cowley} {et~al.}(2002){Cowley}, {Schmidtke}, {Crampton}, \&
  {Hutchings}}]{Cowley2002}
{Cowley}, A.~P., {Schmidtke}, P.~C., {Crampton}, D., \& {Hutchings}, J.~B.
  2002, \aj, 124, 2233

\bibitem[{{Cui} \& {Meng}(2022)}]{Cui2022}
{Cui}, Y. \& {Meng}, X. 2022, \mnras, 515, 2747

\bibitem[{{Cukanovaite} {et~al.}(2019){Cukanovaite}, {Tremblay}, {Freytag},
  {Ludwig}, {Fontaine}, {Brassard}, {Toloza}, \& {Koester}}]{Cukanovaite2019}
{Cukanovaite}, E., {Tremblay}, P.~E., {Freytag}, B., {et~al.} 2019, \mnras,
  490, 1010

\bibitem[{{D'Antona} \& {Tailo}(2020)}]{D'Antona2020}
{D'Antona}, F. \& {Tailo}, M. 2020, arXiv e-prints, arXiv:2011.11385

\bibitem[{{di Stefano} \& {Nelson}(1996)}]{diStefano1996}
{di Stefano}, R. \& {Nelson}, L.~A. 1996, in Supersoft X-Ray Sources, ed.
  J.~{Greiner}, Vol. 472, 3

\bibitem[{{Dilday} {et~al.}(2012){Dilday}, {Howell}, {Cenko}, {Silverman},
  {Nugent}, {Sullivan}, {Ben-Ami}, {Bildsten}, {Bolte}, {Endl}, {Filippenko},
  {Gnat}, {Horesh}, {Hsiao}, {Kasliwal}, {Kirkman}, {Maguire}, {Marcy},
  {Moore}, {Pan}, {Parrent}, {Podsiadlowski}, {Quimby}, {Sternberg}, {Suzuki},
  {Tytler}, {Xu}, {Bloom}, {Gal-Yam}, {Hook}, {Kulkarni}, {Law}, {Ofek},
  {Polishook}, \& {Poznanski}}]{Dilday2012}
{Dilday}, B., {Howell}, D.~A., {Cenko}, S.~B., {et~al.} 2012, Science, 337, 942

\bibitem[{{G{\"a}nsicke} {et~al.}(1998){G{\"a}nsicke}, {van Teeseling},
  {Beuermann}, \& {de Martino}}]{Gansicke1998}
{G{\"a}nsicke}, B.~T., {van Teeseling}, A., {Beuermann}, K., \& {de Martino},
  D. 1998, \aap, 333, 163

\bibitem[{{Greiner} {et~al.}(1991){Greiner}, {Hasinger}, \&
  {Kahabka}}]{Greiner1991}
{Greiner}, J., {Hasinger}, G., \& {Kahabka}, P. 1991, \aap, 246, L17

\bibitem[{{Hachisu} \& {Kato}(2003{\natexlab{a}})}]{Hachisu2003c}
{Hachisu}, I. \& {Kato}, M. 2003{\natexlab{a}}, \apj, 598, 527

\bibitem[{{Hachisu} \& {Kato}(2003{\natexlab{b}})}]{Hachisu2003a}
{Hachisu}, I. \& {Kato}, M. 2003{\natexlab{b}}, \apj, 588, 1003

\bibitem[{{Hachisu} \& {Kato}(2003{\natexlab{c}})}]{Hachisu2003b}
{Hachisu}, I. \& {Kato}, M. 2003{\natexlab{c}}, \apj, 590, 445

\bibitem[{{Hachisu} {et~al.}(1996){Hachisu}, {Kato}, \& {Nomoto}}]{Hachisu1996}
{Hachisu}, I., {Kato}, M., \& {Nomoto}, K. 1996, \apjl, 470, L97

\bibitem[{{Hachisu} {et~al.}(1999{\natexlab{a}}){Hachisu}, {Kato}, \&
  {Nomoto}}]{Hachisu1999b}
{Hachisu}, I., {Kato}, M., \& {Nomoto}, K. 1999{\natexlab{a}}, \apj, 522, 487

\bibitem[{{Hachisu} {et~al.}(1999{\natexlab{b}}){Hachisu}, {Kato}, {Nomoto}, \&
  {Umeda}}]{Hachisu1999a}
{Hachisu}, I., {Kato}, M., {Nomoto}, K., \& {Umeda}, H. 1999{\natexlab{b}},
  \apj, 519, 314

\bibitem[{{Hameury} {et~al.}(1993){Hameury}, {King}, {Lasota}, \&
  {Raison}}]{Hameury1993}
{Hameury}, J.~M., {King}, A.~R., {Lasota}, J.~P., \& {Raison}, F. 1993, \aap,
  277, 81

\bibitem[{{Harpaz} \& {Rappaport}(1991)}]{Harpaz1991}
{Harpaz}, A. \& {Rappaport}, S. 1991, \apj, 383, 739

\bibitem[{{Harpaz} \& {Rappaport}(1994)}]{Harpaz1994}
{Harpaz}, A. \& {Rappaport}, S. 1994, \apj, 434, 283

\bibitem[{{Hillebrandt} \& {Niemeyer}(2000)}]{Hillebrandt2000}
{Hillebrandt}, W. \& {Niemeyer}, J.~C. 2000, \araa, 38, 191

\bibitem[{{Howell}(2011)}]{Howell2011}
{Howell}, D.~A. 2011, Nature Communications, 2, 350

\bibitem[{{Iben} \& {Tutukov}(1984)}]{Iben1984}
{Iben}, I., J. \& {Tutukov}, A.~V. 1984, \apjs, 54, 335

\bibitem[{{Iglesias} \& {Rogers}(1996)}]{Iglesias1996}
{Iglesias}, C.~A. \& {Rogers}, F.~J. 1996, \apj, 464, 943

\bibitem[{{Kahabka} \& {van den Heuvel}(1997)}]{Kahabka1997}
{Kahabka}, P. \& {van den Heuvel}, E.~P.~J. 1997, \araa, 35, 69

\bibitem[{{King} \& {Kolb}(1995)}]{King1995}
{King}, A.~R. \& {Kolb}, U. 1995, \apj, 439, 330

\bibitem[{{Kobayashi} {et~al.}(1998){Kobayashi}, {Tsujimoto}, {Nomoto},
  {Hachisu}, \& {Kato}}]{Kobayashi1998}
{Kobayashi}, C., {Tsujimoto}, T., {Nomoto}, K., {Hachisu}, I., \& {Kato}, M.
  1998, \apjl, 503, L155

\bibitem[{{Langer} {et~al.}(2000){Langer}, {Deutschmann}, {Wellstein}, \&
  {H{\"o}flich}}]{Langer2000}
{Langer}, N., {Deutschmann}, A., {Wellstein}, S., \& {H{\"o}flich}, P. 2000,
  \aap, 362, 1046

\bibitem[{{Liu} \& {Stancliffe}(2018)}]{Liu2018}
{Liu}, Z.-W. \& {Stancliffe}, R.~J. 2018, \mnras, 475, 5257

\bibitem[{{Liu} {et~al.}(2015){Liu}, {Stancliffe}, {Abate}, \&
  {Wang}}]{Liu2015}
{Liu}, Z.-W., {Stancliffe}, R.~J., {Abate}, C., \& {Wang}, B. 2015, \apj, 808,
  138

\bibitem[{{Ma} {et~al.}(2013){Ma}, {Chen}, {Chen}, {Denissenkov}, \&
  {Han}}]{Ma2013}
{Ma}, X., {Chen}, X., {Chen}, H.-l., {Denissenkov}, P.~A., \& {Han}, Z. 2013,
  \apjl, 778, L32

\bibitem[{{Maoz} {et~al.}(2014){Maoz}, {Mannucci}, \& {Nelemans}}]{Maoz2014}
{Maoz}, D., {Mannucci}, F., \& {Nelemans}, G. 2014, \araa, 52, 107

\bibitem[{{Meng} {et~al.}(2009){Meng}, {Chen}, \& {Han}}]{Meng2009a}
{Meng}, X., {Chen}, X., \& {Han}, Z. 2009, \mnras, 395, 2103

\bibitem[{{Meng} \& {Podsiadlowski}(2017)}]{Meng2017}
{Meng}, X. \& {Podsiadlowski}, P. 2017, \mnras, 469, 4763

\bibitem[{{Meng} \& {Yang}(2010)}]{Meng2010}
{Meng}, X. \& {Yang}, W. 2010, \apj, 710, 1310

\bibitem[{{Meyer-Hofmeister} {et~al.}(1997){Meyer-Hofmeister}, {Schandl}, \&
  {Meyer}}]{Meyer-Hofmeister1997}
{Meyer-Hofmeister}, E., {Schandl}, S., \& {Meyer}, F. 1997, \aap, 321, 245

\bibitem[{{Nomoto} {et~al.}(1979){Nomoto}, {Nariai}, \&
  {Sugimoto}}]{Nomoto1979}
{Nomoto}, K., {Nariai}, K., \& {Sugimoto}, D. 1979, \pasj, 31, 287

\bibitem[{{Nomoto} {et~al.}(2007){Nomoto}, {Saio}, {Kato}, \&
  {Hachisu}}]{Nomoto2007}
{Nomoto}, K., {Saio}, H., {Kato}, M., \& {Hachisu}, I. 2007, \apj, 663, 1269

\bibitem[{{Nomoto} {et~al.}(1984){Nomoto}, {Thielemann}, \&
  {Yokoi}}]{Nomoto1984}
{Nomoto}, K., {Thielemann}, F.~K., \& {Yokoi}, K. 1984, \apj, 286, 644

\bibitem[{{Pakull} {et~al.}(1993){Pakull}, {Motch}, {Bianchi}, {Thomas},
  {Guibert}, {Beaulieu}, {Grison}, \& {Schaeidt}}]{Pakull1993}
{Pakull}, M.~W., {Motch}, C., {Bianchi}, L., {et~al.} 1993, \aap, 278, L39

\bibitem[{{Patat} {et~al.}(2007){Patat}, {Chandra}, {Chevalier}, {Justham},
  {Podsiadlowski}, {Wolf}, {Gal-Yam}, {Pasquini}, {Crawford}, {Mazzali},
  {Pauldrach}, {Nomoto}, {Benetti}, {Cappellaro}, {Elias-Rosa}, {Hillebrandt},
  {Leonard}, {Pastorello}, {Renzini}, {Sabbadin}, {Simon}, \&
  {Turatto}}]{Patat2007}
{Patat}, F., {Chandra}, P., {Chevalier}, R., {et~al.} 2007, Science, 317, 924

\bibitem[{{Paxton} {et~al.}(2011){Paxton}, {Bildsten}, {Dotter}, {Herwig},
  {Lesaffre}, \& {Timmes}}]{Paxton2011}
{Paxton}, B., {Bildsten}, L., {Dotter}, A., {et~al.} 2011, \apjs, 192, 3

\bibitem[{{Paxton} {et~al.}(2013){Paxton}, {Cantiello}, {Arras}, {Bildsten},
  {Brown}, {Dotter}, {Mankovich}, {Montgomery}, {Stello}, {Timmes}, \&
  {Townsend}}]{Paxton2013}
{Paxton}, B., {Cantiello}, M., {Arras}, P., {et~al.} 2013, \apjs, 208, 4

\bibitem[{{Paxton} {et~al.}(2015){Paxton}, {Marchant}, {Schwab}, {Bauer},
  {Bildsten}, {Cantiello}, {Dessart}, {Farmer}, {Hu}, {Langer}, {Townsend},
  {Townsley}, \& {Timmes}}]{Paxton2015}
{Paxton}, B., {Marchant}, P., {Schwab}, J., {et~al.} 2015, \apjs, 220, 15

\bibitem[{{Paxton} {et~al.}(2018){Paxton}, {Schwab}, {Bauer}, {Bildsten},
  {Blinnikov}, {Duffell}, {Farmer}, {Goldberg}, {Marchant}, {Sorokina},
  {Thoul}, {Townsend}, \& {Timmes}}]{Paxton2018}
{Paxton}, B., {Schwab}, J., {Bauer}, E.~B., {et~al.} 2018, \apjs, 234, 34

\bibitem[{{Paxton} {et~al.}(2019){Paxton}, {Smolec}, {Schwab}, {Gautschy},
  {Bildsten}, {Cantiello}, {Dotter}, {Farmer}, {Goldberg}, {Jermyn}, {Kanbur},
  {Marchant}, {Thoul}, {Townsend}, {Wolf}, {Zhang}, \& {Timmes}}]{Paxton2019}
{Paxton}, B., {Smolec}, R., {Schwab}, J., {et~al.} 2019, \apjs, 243, 10

\bibitem[{{Perlmutter} {et~al.}(1999){Perlmutter}, {Aldering}, {Goldhaber},
  {Knop}, {Nugent}, {Castro}, {Deustua}, {Fabbro}, {Goobar}, {Groom}, {Hook},
  {Kim}, {Kim}, {Lee}, {Nunes}, {Pain}, {Pennypacker}, {Quimby}, {Lidman},
  {Ellis}, {Irwin}, {McMahon}, {Ruiz-Lapuente}, {Walton}, {Schaefer}, {Boyle},
  {Filippenko}, {Matheson}, {Fruchter}, {Panagia}, {Newberg}, {Couch}, \&
  {Project}}]{Perlmutter1999}
{Perlmutter}, S., {Aldering}, G., {Goldhaber}, G., {et~al.} 1999, \apj, 517,
  565

\bibitem[{{Podsiadlowski}(1991)}]{PH.1991}
{Podsiadlowski}, P. 1991, \nat, 350, 136

\bibitem[{{Pols} {et~al.}(1998){Pols}, {Schr{\"o}der}, {Hurley}, {Tout}, \&
  {Eggleton}}]{Pols1998}
{Pols}, O.~R., {Schr{\"o}der}, K.-P., {Hurley}, J.~R., {Tout}, C.~A., \&
  {Eggleton}, P.~P. 1998, \mnras, 298, 525

\bibitem[{{Rappaport} {et~al.}(1995){Rappaport}, {Podsiadlowski}, {Joss}, {Di
  Stefano}, \& {Han}}]{Rappaport1995}
{Rappaport}, S., {Podsiadlowski}, P., {Joss}, P.~C., {Di Stefano}, R., \&
  {Han}, Z. 1995, \mnras, 273, 731

\bibitem[{{Reinsch} {et~al.}(1996){Reinsch}, {van Teeseling}, {Beuermann}, \&
  {Abbott}}]{Reinsch1996}
{Reinsch}, K., {van Teeseling}, A., {Beuermann}, K., \& {Abbott}, T.~M.~C.
  1996, \aap, 309, L11

\bibitem[{{Reinsch} {et~al.}(2000){Reinsch}, {van Teeseling}, {King}, \&
  {Beuermann}}]{Reinsch2000}
{Reinsch}, K., {van Teeseling}, A., {King}, A.~R., \& {Beuermann}, K. 2000,
  \aap, 354, L37

\bibitem[{{Riess} {et~al.}(1998){Riess}, {Filippenko}, {Challis},
  {Clocchiatti}, {Diercks}, {Garnavich}, {Gilliland}, {Hogan}, {Jha},
  {Kirshner}, {Leibundgut}, {Phillips}, {Reiss}, {Schmidt}, {Schommer},
  {Smith}, {Spyromilio}, {Stubbs}, {Suntzeff}, \& {Tonry}}]{Riess1998}
{Riess}, A.~G., {Filippenko}, A.~V., {Challis}, P., {et~al.} 1998, \aj, 116,
  1009

\bibitem[{{Ritter} {et~al.}(1994){Ritter}, {Zhang}, \& {Kolb}}]{Ritter1994}
{Ritter}, H., {Zhang}, Z.~Y., \& {Kolb}, U. 1994, in Astronomische Gesellschaft
  Abstract Series, Vol.~10, Astronomische Gesellschaft Abstract Series, 66

\bibitem[{{Ritter} {et~al.}(2000){Ritter}, {Zhang}, \& {Kolb}}]{Ritter2000}
{Ritter}, H., {Zhang}, Z.~Y., \& {Kolb}, U. 2000, \aap, 360, 959

\bibitem[{{Rodney} {et~al.}(2015){Rodney}, {Riess}, {Scolnic}, {Jones},
  {Hemmati}, {Molino}, {McCully}, {Mobasher}, {Strolger}, {Graur}, {Hayden}, \&
  {Casertano}}]{Rodney2015}
{Rodney}, S.~A., {Riess}, A.~G., {Scolnic}, D.~M., {et~al.} 2015, \aj, 150, 156

\bibitem[{{Schaeidt} {et~al.}(1993){Schaeidt}, {Hasinger}, \&
  {Truemper}}]{Schaeidt1993}
{Schaeidt}, S., {Hasinger}, G., \& {Truemper}, J. 1993, \aap, 270, L9

\bibitem[{{Southwell} {et~al.}(1996){Southwell}, {Livio}, {Charles},
  {O'Donoghue}, \& {Sutherland}}]{Southwell1996}
{Southwell}, K.~A., {Livio}, M., {Charles}, P.~A., {O'Donoghue}, D., \&
  {Sutherland}, W.~J. 1996, \apj, 470, 1065

\bibitem[{{Sternberg} {et~al.}(2011){Sternberg}, {Gal-Yam}, {Simon}, {Leonard},
  {Quimby}, {Phillips}, {Morrell}, {Thompson}, {Ivans}, {Marshall},
  {Filippenko}, {Marcy}, {Bloom}, {Patat}, {Foley}, {Yong}, {Penprase},
  {Beeler}, {Allende Prieto}, \& {Stringfellow}}]{Sternberg2011}
{Sternberg}, A., {Gal-Yam}, A., {Simon}, J.~D., {et~al.} 2011, Science, 333,
  856

\bibitem[{{Tremblay} {et~al.}(2015){Tremblay}, {Ludwig}, {Freytag}, {Fontaine},
  {Steffen}, \& {Brassard}}]{Tremblay2015}
{Tremblay}, P.~E., {Ludwig}, H.~G., {Freytag}, B., {et~al.} 2015, \apj, 799,
  142

\bibitem[{{van den Heuvel} {et~al.}(1992){van den Heuvel}, {Bhattacharya},
  {Nomoto}, \& {Rappaport}}]{vandenHeuvel1992}
{van den Heuvel}, E.~P.~J., {Bhattacharya}, D., {Nomoto}, K., \& {Rappaport},
  S.~A. 1992, \aap, 262, 97

\bibitem[{{van Teeseling} {et~al.}(1996){van Teeseling}, {Beuermann}, \&
  {Verbunt}}]{vanTeeseling1996}
{van Teeseling}, A., {Beuermann}, K., \& {Verbunt}, F. 1996, \aap, 315, 467

\bibitem[{{Webbink}(1984)}]{Webbink1984}
{Webbink}, R.~F. 1984, \apj, 277, 355

\bibitem[{{Whelan} \& {Iben}(1973{\natexlab{a}})}]{Whelan1973}
{Whelan}, J. \& {Iben}, Icko, J. 1973{\natexlab{a}}, \apj, 186, 1007

\bibitem[{{Whelan} \& {Iben}(1973{\natexlab{b}})}]{Whenlan1973}
{Whelan}, J. \& {Iben}, Icko, J. 1973{\natexlab{b}}, \apj, 186, 1007

\bibitem[{{Wolf} {et~al.}(2013){Wolf}, {Bildsten}, {Brooks}, \&
  {Paxton}}]{Wolf2013}
{Wolf}, W.~M., {Bildsten}, L., {Brooks}, J., \& {Paxton}, B. 2013, \apj, 777,
  136

\end{thebibliography}
	\bibliographystyle{aa} 

\end{document}